\documentclass[onecolumn,aps,prd,superscriptaddress,nofootinbib,amsmath,amssymb,floats,floatfix,showkeys,notitlepage,longbibliography]{revtex4-2}
\usepackage[dvips]{graphicx,color}
\usepackage{subfigure}
\usepackage{multirow}
\usepackage{times}
\usepackage{float}
\usepackage{tabularx}
\usepackage{makecell}
\usepackage{booktabs,subcaption,amsfonts,dcolumn}
\newcolumntype{d}[1]{D..{#1}}
\usepackage{array}
\usepackage{mathrsfs}  
\usepackage{bm}
\usepackage{xcolor}
\usepackage[%
  colorlinks=true,
  urlcolor=blue,
  linkcolor=red,
  citecolor=blue
]{hyperref}
\usepackage{orcidlink}

\newcommand\emtensor{\mathrel{\stackrel{\makebox[0pt]{\mbox{\normalfont\tiny e-m}}}{\mathcal{T}}}}
\newcommand\seteq{\mathrel{\stackrel{\makebox[0pt]{\mbox{\normalfont\tiny set}}}{=}}}

\begin{document}
\title{Influence of GUP corrected Casimir energy on zero tidal force wormholes in modified teleparallel gravity with matter coupling}

\author{Mohammed Muzakkir Rizwan \orcidlink{0009-0008-6375-9045}}
\email{mohammed.muzu.riz@gmail.com}
\affiliation{Department of Mathematics, Birla Institute of Technology and
Science-Pilani,\\ Hyderabad Campus, Hyderabad-500078, India.}

\author{Zinnat Hassan \orcidlink{0000-0002-3223-4085}}
\email{zinnathassan980@gmail.com}
\affiliation{Department of Mathematics, Birla Institute of Technology and
Science-Pilani,\\ Hyderabad Campus, Hyderabad-500078, India.}

\author{P.K. Sahoo \orcidlink{0000-0003-2130-8832}}
\email{pksahoo@hyderabad.bits-pilani.ac.in}
 \affiliation{Department of Mathematics, Birla Institute of Technology and
 Science-Pilani,\\ Hyderabad Campus, Hyderabad-500078, India.}

\author{Ali \"Ovg\"un \orcidlink{0000-0002-9889-342X}}
\email{ali.ovgun@emu.edu.tr}
\affiliation{Physics Department, Eastern Mediterranean
University, Famagusta, 99628 North Cyprus, via Mersin 10, Turkiye}
\date{\today}

\begin{abstract}
In recent times, the study of the Casimir effect in quantum field theory has garnered increasing attention because of its potential to be an ideal source of exotic matter needed for stabilizing traversable wormholes. It has been confirmed through experimental evidence that this phenomenon involves fluctuations in the vacuum field, leading to a negative energy density. Motivated by the above, we have investigated Casimir wormholes with corrections from the Generalized Uncertainty Principle (GUP) within the framework of matter-coupled teleparallel gravity. Our analysis includes three well-known GUP models: the Kempf, Mangano, and Mann (KMM) model, the Detournay, Gabriel, and Spindel (DGS) model, and a third model called Model II. For a broader analysis, we have considered two well-known model functions for the teleparallel theory: a linear $f(T,\mathcal{T})=\alpha T+\beta\mathcal{T}$ and a quadratic model $f(T,\mathcal{T})=\eta T^2+\chi\mathcal{T}$. The shape function solutions corresponding to both models are examined in the absence of tidal forces in spacetime. We also demonstrate the crucial role played by the parameters of the $f(T,\mathcal{T})$ models in the violation of the energy conditions. With the increasing interest in detecting gravitational waves from astrophysical objects, we have thoroughly discussed the perturbation of the wormhole solutions in the scalar, electromagnetic, axial gravitational, and Dirac field backgrounds. We employ the $3^{rd}$ order WKB expansion to find the complex frequencies associated with the quasinormal modes of energy dissipation. Additionally, we also calculate the active mass and total gravitational energy for the wormhole geometry. The amount of exotic matter involved in sustaining these wormholes is also found in this paper. Furthermore, the physical stability of such Casimir wormholes is examined using the Tolman-Oppenheimer-Volkoff equation.
\end{abstract}

\maketitle
\section{Introduction}
Wormholes are hypothetical bridge-like connections between two distinct spacetimes, potentially allowing shortcuts for interstellar travel. Although wormholes have yet to be detected, the debates around their existence and how they might form are still unresolved. The possibility of creating a traversable wormhole or fabricating one in a laboratory in the far future motivates theoretical researchers, drawing significant interest recently. Regardless, recent progress in quantum mechanics suggests that real-world traversable wormholes might exist, raising intriguing questions about their nature and potential uses. Some researchers have even proposed that they could be created in a laboratory and used in quantum teleportation experiments \cite{PRXQuantum.4.010320,PRXQuantum.4.010321}. Recently, an interesting analysis of the light ring (LR) structure of a general wormhole spacetime was investigated in \cite{PhysRevD.109.124065}. Moreover, wormhole solutions in the galactic halo region were studied in \cite{PhysRevD.109.064043,Mehedi2022,https://doi.org/10.1002/prop.202200129,Mustafa_2022}. The thin shell around wormholes via solitonic quantum wave dark matter has been explored in \cite{MUSTAFA2024169551}. Worldsheet traversable wormholes \cite{12} and the existence of humanly traversable wormholes \cite{PhysRevD.103.066007} have also been discussed in the literature. Recently, wormhole shadows and gravitational lensing have been investigated using the Karmarkar condition, as reported in \cite{Molla2024}. Further, wormholes in the braneworld cosmology, connecting branes across the higher dimensional bulk, have also made waves in the current pursuit of quantum gravity \cite{bronnikov2003possible,wang2018traversable,biswas2022echoes,Sengupta_2022}.
Furthermore, the deflection of light by wormholes was discussed in \cite{14,PhysRevD.101.044001,PhysRevD.109.043032,https://doi.org/10.1002/andp.202300039} and wormhole shadow in \cite{PhysRevD.107.104060,PhysRevD.109.104012,13,Rahaman_2021}. One may go through the following intriguing works on wormhole geometries in literature \cite{PhysRevLett.128.091104,PhysRevD.106.044035,Maldacena_2023,Araujo_Filho_2023,Cruz_2024,CLEMENT2023137677,PhysRevD.109.103034,PhysRevD.109.064007,PhysRevD.108.124049,MUSTAFA202432}.

In 1916, Flamm first proposed the theoretical framework of a wormhole \cite{flamm1916beitrage} and realized that the Schwarzschild black hole might offer a means for interstellar journeys. Later, Einstein and Rosen \cite{einsteinrosen1935} expounded on the works of Flamm and constructed a spacetime bridge known as the Einstein-Rosen Bridge. 
As stunning as it may sound, the wormhole solutions were proven to be highly unstable, given how its throat closes immediately even for the slightest perturbation \cite{kruskal1960maximal,fuller1962causality,eardley1974death}. Later, in 1988, Morris and Thorne first proposed the concept of a traversable wormhole \cite{morris1988wormholes} by theorizing certain desired properties the wormhole must possess for its traversability. In particular, a traversable wormhole must possess exotic matter at/near its throat to disobey the energy conditions \cite{morris1988wormholes,visser1995lorentzian}, specifically the null energy condition (NEC), which states: $\emtensor_{\mu\nu}k^\mu k^\nu<0$, where $\emtensor_{\mu\nu}$ is the energy-momentum tensor and $k^\mu,\,k^\nu$ are null vectors.

It is widely accepted that General Relativity (GR) effectively describes the observable universe. However, some recent observations have prompted the need to modify classical relativity to explain certain phenomena, such as the late-time acceleration expansion of the universe \cite{Perlmutter_1999,Komatsu_2011,Riess_2007}. As a result, various modified theories have been proposed to address these issues, including $f(R)$ gravity \cite{STAROBINSKY198099}, $f(R, T)$ gravity \cite{PhysRevD.84.024020}, $f(Q)$ gravity \cite{PhysRevD.98.044048}, and others. These modified theories of gravity offer potential insights into cosmic expansion and related concepts, making them fascinating candidates for exploring astrophysical objects like wormholes. Rastgoo and Parsaei \cite{Parsaei2024} investigated wormhole solutions in $f(R,T)$ gravity with variable linear equations of state (E0S), revealing that their solutions meet all energy conditions. In \cite{KIRORIWAL2024101559}, the authors discussed wormhole solutions by considering a particular EoS in the framework of $f(Q)$ gravity and showed the violation of NEC near the throat. Malik et al. \cite{Malik_2022} studied traversable wormholes in $f(R)$ gravity by applying the Karmarkar condition to various redshift functions. In \cite{Fayyaz2020}, wormhole solutions under different $f(R)$ models were analyzed using this condition. In $f(R,T)$ gravity, \cite{ERREHYMY2024102972} demonstrated that wormholes can satisfy energy conditions with karmarkar conditions. Further, the stability of the thin shell around wormholes in the context of  $f(Q)$ gravity \cite{https://doi.org/10.1002/prop.202200053}, Rastall gravity \cite{Ali2021} and $4D$ EGB gravity \cite{GODANI2022100952} were also discussed. Further, Mustafa et al. \cite{MUSTAFA2024101508} studied the effect of cold dark matter and solitonic quantum wave dark matter on wormhole geometry in the context of $f(R,L_m)$ gravity. Furthermore, the effect of redshift functions on weak energy conditions in $f(R)$ gravity has been studied in \cite{Sadeghi2024}. For more interesting works on wormhole geometry, readers can check some recent articles within the framework of modified theories of gravity, including $f(R)$ gravity \cite{shamir2020traversable,EPJCDeFalco,MUNIZ2022169129,Gogoi_2023,PhysRevD.108.024063,Abdelghani2024}, $f(R,T)$ gravity \cite{SARKAR2024101439,Luis_Rosa,Lu_Jianbo,Fatima2023}, $f(Q)$ gravity \cite{Lu_Jianbo2024,Tayde_Moreshwar,Rastgoo_S,CHAUDHARY2024103002,PRADHAN2024101620}, and other modified theories of gravity \cite{Qadeer2024,Luís_Rosa,Ilyas_2023,PhysRevD.108.104008,Nguyen2023,TAYDE2023101288,Hassan_Zinnat,https://doi.org/10.1002/andp.202400155}.\\
\indent To further envelope this idea, the torsion ($T$, torsion scalar) of space-time was used as an object of importance over its curvature \cite{capozziello2011cosmography, krvsvsak2019teleparallel}. $f(T)$ gravity, also known as the teleparallel theory of gravity, was first used by Einstein to develop a unified theory of gravity and electromagnetism \cite{einstein2005riemann}. The Teleparallel formalism utilizes a curvature-free linear connection defined in terms of the tetrad field, and there are interesting articles on wormholes using this formalism in \cite{boehmer2012wormhole, jamil2013wormholes, sharif2013wormhole}. 
Additionally, T. Harko et al. \cite{harko2014f} have presented an MTG that formalizes the geometry-matter coupling to teleparallelism, i.e., the $f(T, \mathcal{T})$ formalism. Teleparallel theories of gravity have proven to be compatible with observed cosmological data, such as the Supernovae type Ia (SNIa), the Baryonic Acoustic Oscillation (BAO), and the Cosmic Microwave Background (CMB) radiation \cite{wu2010observational}, with additional constraints using cosmic chronometer method and gravitational wave astronomy \cite{nunes2016new,nunes2018new}. Researchers have also studied late-time cosmic expansion using the dynamics of teleparallel dark energy in both quintessence-like and phantom-like regions \cite{geng2011teleparallel,geng2012observational,wei2012dynamics,gu2013teleparallel}, which could potentially replace the $\Lambda$CDM model. Further, there are works on Gravitational Waves (GWs) in the weak field limit \cite{farrugia2018gravitational,bamba2013no}. Concerning the matter-coupled formalism, \cite{arora2023squared} have used the squared-torsion model in $f(T, \mathcal{T})$ gravity and compared it with the observational data. Ditta et al. \cite{ditta2021study} have studied the wormhole solution using the Karmarkar condition in a linear and non-linear $f(T, \mathcal{T})$ model, while Mustafa et al. \cite{mustafa2021traversable} have studied the wormhole geometry using the Gaussian and Lorentzian densities as a source for exotic matter in a linear $f(T, \mathcal{T})$ framework.

In the context of GR, it is widely recognized that the NEC must be violated for a wormhole to be traversable, indicating the essential of exotic matter at the wormhole's throat. One practical example of such matters can be found in the Casimir effect. The Casimir effect occurs when two parallel conductive plates are placed in a vacuum, leading to their attraction due to the energy generation between them by the zero-point fluctuations of quantum field theory. This effect was initially identified in literature referenced as \cite{Polder1948} and further elaborated on through a different method in \cite{,Dzyaloshinskii_1961}. The experimental verification of the Casimir effect has been reported in \cite{PhysRevLett.78.5,PhysRevLett.88.041804}. Recently, in \cite{Garattini2019}, the author introduced a wormhole solution exploring the negative energy density resulting from the Casimir effect and examined the importance of the quantum weak energy conditions on the traversability of the wormhole.

The motive for the Generalized Uncertainty Principle (GUP) comes from the fact that in quantum gravity, there is a fundamental length scale beyond which further resolution is not possible, such as in string theory and the length of string, etc. Moreover, introducing a minimum length scale presents various significant phenomenological effects within quantum gravity, thoroughly discussed in \cite{Hossenfelder2013}. In quantum mechanics, the uncertainty principle involving two Hermitian operators (observables), say $\hat{A}$ and $\hat{B}$, can be defined as
\begin{equation}
    \Delta\hat{A}\Delta\hat{B}\geq \frac{1}{2i}\langle[\hat{A}, \hat{B}]\rangle,
\end{equation}
where $\Delta\hat{A}=\sqrt{\langle\hat{A}^2\rangle-(\langle\hat{A}\rangle)^2}$.
Hence, the usual Heisenberg's Uncertainty Principle (HUP) for the position ($\hat{x}$) and momentum ($\hat{p}$) operators follows as $\Delta\hat{x}\Delta\hat{p}\geq\frac{\hbar}{2}$, where $\hbar$ is the reduced Planck constant. Despite the strong experimental evidence backing up HUP, we run into problems when we incorporate it into GR, we get a singularity in the metric like in the Schwarzschild solutions at $r=0$.\\
\indent It is apparent from the HUP that the position and momentum near the singularity would behave like $\Delta x\sim\frac{const.}{\Delta p}$ but the Schwarzschild solutions go $\Delta x=2G\Delta p$, in natural units. We can incorporate HUP with the Schwarzchild singularity by the following ansatz for position-momentum relation: $\Delta x\sim\frac{\hbar}{\Delta p}+G\Delta p$. We can see that the commutation relation roughly becomes $[x,p]=i\hbar(1+\lambda p^2)$, and the fundamental scale goes $\Delta p_m\sim\sqrt{\hbar/G}$ and $\Delta x_m\sim\sqrt{\hbar G}$ in the natural units. Studies have proposed various experiments to observe the modified dispersion relation of a photon through gamma-ray bursts \cite{Amelino-Camelia}. There is a discussion in \cite{PhysRevLett.101.221301} on how different experimental methods can probe the fundamental length scale ($\lambda$), such as the lamb shift $(\lambda < 10^{36})$, Landau level $(\lambda < 10^{56})$, and tunneling $(\lambda < 10^{21})$. For more experiments, one can refer to some Refs. \cite{Hossenfelder2013,universe8030192}. The Casimir effect in minimal-length theories based on GUP corrections was discussed in \cite{PhysRevD.85.045030}. The application of minimal time scale and GUP has been successfully used to solve the Wheeler-Dewitt equation for the universe's evolution \cite{doi:10.1142/S0217751X1750049X}. Further,
GUP has also been used to find dispersion relations during Hawking radiation of Schwarzschild-de Sitter black holes, and one can also get a limit of minimal length scale already knowing the black hole evaporation formula using semiclassical quantum gravity \cite{arraut2009comparing}.
Studies have explored the formalism of Casimir wormholes, based on different models in extended gravity theories and various physical factors ranging from differing geometrical structures of the Casimir plates to the temperature-based perturbations to the wormhole throat \cite{azmat2024class,hassan2022casimir,hassan2023gup,muniz2021casimir,garattini2024hot}.

\indent This paper thoroughly investigates traversable wormholes in $f(T,\mathcal{T})$ formalism, that source the required exotic matter from the Casimir effect with GUP corrections that arise from the minimal length concept in quantum mechanics. The article is structured as follows: We discuss the mathematical foundations for the Casimir effect, followed by the GUP in Sec.\ref{sec-CasimirGUP}. In Sec.\ref{sec-modgrav}, we introduce the $f(T, \mathcal{T})$ gravity theory and the resulting expressions for energy conditions for the anisotropic matter fluid. We consider the linear $f(T,\mathcal{T})=\alpha T+\beta\mathcal{T}$ and non-linear $f(T,\mathcal{T})=\eta T^2+\chi\mathcal{T}$ models and find the wormhole shape function solution in absence of gravitational tides in Sec.\ref{sec-Solution}. In the following Sec.\ref{sec-QNMs}, the Quasinormal modes (QNMs) for the above wormhole geometries are found for the scalar, vector (electromagnetic), tensor (axial gravitational), and Dirac field perturbations. Additionally, we investigate essential physical attributes of a wormhole, such as the active mass profiles, the total gravitational energy possessed by the wormhole, the amount of exotic matter for the wormhole's maintenance, stability of the wormhole structure in Sec.\ref{sec-physicalprops}. Finally, we conclude with our paper in the last section.\\
\indent In this analysis, we consider the following choice of natural units: $c = \hbar = 1$.

\section{Casimir Energy and GUP corrections}\label{sec-CasimirGUP}
This section will briefly discuss the fundamentals of the Casimir effect and the motivation for the Generalized Uncertainty Principle (GUP) based corrections to the Casimir energy. In the following sections, we will apply these models to wormhole geometry to observe their effects.
\subsection{Casimir Effect}
One of the potential natural sources of exotic matter is Casimir energy. The Casimir effect is the phenomenon in QFT where, when two uncharged metallic plane plates are held parallel to one another at an extremely close ($\sim$ a micron apart) distance in a vacuum, a force comes into being that pushes them towards each other. This is caused by the zero-point energy of the quantum electrodynamics being distorted by the plates, which has some bearing on the negative Casimir energy. It was theoretically proposed in 1948 \cite{casimir1948attraction} and independently studied by \cite{dzyaloshinskii1961general}, later experimentally verified in the Philips laboratory \cite{black1960measurements}. However, it was only in recent years that more reliable experimental investigations were employed to confirm this phenomenon \cite{lamoreaux1997demonstration,bressi2002measurement}.\\
The renormalized negative energy between the conducting plates is mathematically expressed as
\begin{equation}\label{casimirenergy}
    E(a)=-\frac{\hbar c\pi^2S}{720a^3},
\end{equation}
where `$a$' is the separation between the parallel plates and `$S$' is their surface area.\\
One can derive the expression (\ref{casimirenergy}) by summing over the normal modes of the field and adequately regularizing the sum. The regularization can be done in two ways: first, by introducing a cut-off limit \cite{zee2010quantum}, and second via an analytic continuation of the Riemann Zeta function \cite{padmanabhan2016quantum}, both yielding the same result.\\
Consequently, energy density can be found by dividing (\ref{casimirenergy}) with the volume between the plates $V=aS$
\begin{equation}\label{casimirdensity}
    \rho_c=\frac{E}{aS}=-\frac{\hbar c\pi^2}{720a^4}
\end{equation}
The pressure on the surface of the plates is
\begin{equation}
    p_c=\frac{F_c}{S}=-\frac{1}{S}\frac{dE}{da}=-3\frac{\hbar c\pi^2}{720a^4}.
\end{equation}
Thus, obeying the linear EoS, $\rho_c=\omega p_c$, where $\omega=3$.
\subsection{Generalized Uncertainty Principle}
The GUP formulation improves on the well-known HUP, which seeks to incorporate quantum gravity effects at minimal length. The existence of a minimal length in the theory explicitly limits the resolution of small distances in spacetime. This also leads to modifying the uncertainty principle, which in one dimension looks like 
\begin{equation}
    \Delta x\Delta p\geq\frac{\hbar}{2}[1+\lambda(\Delta p)^2]
\end{equation}
expanded up to the first order in the minimal length parameter $\lambda$. This new uncertainty relation implies the following modified commutation relation in the first order: $[\hat{x},\hat{p}]=i\hbar(1+\lambda \hat{p}^2)$.

We also note that position and momentum are no longer conjugate variables in the usual sense, hence, in these theories the position eigenstates can no longer be interpreted as actual physical positions. However, we can still talk about the positions as the states projected onto the set of maximally localized states, called the ``quasi-position representation" as detailed in \cite{frassino2012casimir}. These maximally localized states $|\psi^{ML}_x\rangle$ minimize the uncertainty in position $(\Delta x)_{|\psi^{ML}_x\rangle}=\Delta x_0$ and are centered around some average position $\langle\psi^{ML}_x|\hat{x}|\psi^{ML}_x\rangle=x$.\\
From \cite{frassino2012casimir}, the generalized commutation relation in $n$ spatial dimensions is
\begin{equation}
    [\hat{x}_i,\hat{p}_j]=i\hbar\big[f(\hat{p}^2)\delta_{ij}+g(\hat{p}^2)\hat{p}_i\hat{p}_j\big]\;\;\;\;\;\;i,j=1,2,\ldots,n
\end{equation}
where $f(\hat{p}^2)$ and $g(\hat{p}^2)$ are generic functions. These functions are not completely arbitrary and can be obtained by imposing the translational and rotational invariance on the generalized commutation relation. We observe that these are the only options in the first order because of spherical symmetry. The general form of maximally localized states around an average position `$\bm{r}$' in the momentum representation is given by
\begin{equation}\label{ML states}
    \langle p|\psi^{ML}_{\bm{r}}\rangle=\frac{1}{(2\pi\hbar)^{3/2}}\Omega(p)\,exp\Biggl\{-\frac{i}{\hbar}\left[\bm{\kappa}(p)\cdot \bm{r}-\hbar\,\omega(p)\,t\right]\Biggl\},
\end{equation}
where $p=|\bm{\hat{p}}|$, $\omega(p)$ represents the dispersion relation, $\Omega(p)$ denotes the measure, and $\bm{\kappa}(p)$ is the wave vector. Furthermore, the above wave-function satisfies the equation
\begin{equation}
    \Biggr[\hat{x}-\langle \hat{x}\rangle+\frac{\langle[\hat{x},\hat{p}]\rangle}{2(\Delta p)^2}(\hat{p}-\langle \hat{p}\rangle)\Biggr]|\psi\rangle=0
\end{equation}
because these states minimize the uncertainty in position, as discussed earlier \cite{kempf1995hilbert}. In this paper, we consider two different models that have been extensively discussed in literature \cite{kempf1995hilbert,panella2007casimir}, one of them has been studied within two different approaches: KMM \cite{kempf1995hilbert} which utilizes squeeze state, and DGS \cite{detournay2002maximally} which uses the variational principle approach at determining the maximally localized states.

\subsubsection{KMM model}
This model corresponds to the choice of the generic functions $f(\hat{p}^2)$ and $g(\hat{p}^2)$ given in \cite{kempf1997minimal}
\begin{equation}\label{KMM generic}
    f(\hat{p}^2)=\frac{\lambda \hat{p}^2}{\sqrt{1+2\lambda\hat{p}^2}-1},\;\;\;\;\;g(\hat{p}^2)=\lambda.
\end{equation}
From this point onwards, we will drop the hat over the operators, by virtue of it being self-explanatory. The KMM construction assigns the following functions to the maximally localized states (\ref{ML states}):
\begin{equation}
    \kappa_i(p)=\left(\frac{\sqrt{1+2\lambda p^2}-1}{\lambda p^2}\right)p_i,\;\;\;\;\;\omega(p)=\frac{pc}{\hbar}\left(\frac{\sqrt{1+2\lambda p^2}-1}{\lambda p^2}\right),
\end{equation}
\begin{equation}
    \Omega(p)=\left(\frac{\sqrt{1+2\lambda p^2}-1}{\lambda p^2}\right)^{\alpha/2},
\end{equation}
where $n$ is the number of spatial dimensions and $\alpha=1+\sqrt{1+n/2}$ is a numerical constant characteristic of the KMM approach. Using the scalar product of the maximally localized states, one can represent the identity operator as:
\begin{equation}
    \int \frac{d^np}{\sqrt{1+2\lambda p^2}}\left(\frac{\sqrt{1+2\lambda p^2}-1}{\lambda p^2}\right)^{n+\alpha}|\bm{p}\rangle\langle\bm{p}|=\bm{1}.
\end{equation}
\subsubsection{DGS model}
As formerly discussed, distinct maximally localized states may correspond to a given choice of generic functions (\ref{KMM generic}). The maximally localized states using DGS construction are given by (\ref{ML states}) with:
\begin{equation}
    \kappa_i(p)=\left(\frac{\sqrt{1+2\lambda p^2}-1}{\lambda p^2}\right)p_i,\;\;\;\;\;\omega(p)=\frac{pc}{\hbar}\left(\frac{\sqrt{1+2\lambda p^2}-1}{\lambda p^2}\right),
\end{equation}
\begin{align}
    \Omega(p)&=\Biggl[\Gamma\left(\frac{3}{2}\right)\left(\frac{2\sqrt{2}}{\pi\sqrt{\lambda}}\right)^{1/2}\Biggr]\left(\frac{1}{p}\frac{\lambda p^2}{\sqrt{1+2\lambda p^2}-1}\right)^{1/2}\bm{J}_{\frac{1}{2}}\Biggl[\frac{\pi\sqrt{\lambda}}{\sqrt{2}}\left(\frac{\sqrt{1+2\lambda p^2}-1}{\lambda p^2}\right)p\Biggr]\\
    &\notag=\frac{\sqrt{2}}{\pi}\frac{\sqrt{\lambda p^2}}{\sqrt{1+2\lambda p^2}-1}sin\left(\frac{\sqrt{2}\pi(\sqrt{1+2\lambda p^2}-1)}{2\sqrt{\lambda p^2}}\right),
\end{align}
where $\bm{J}_{\frac{1}{2}}$ is the Bessel function of the first kind. The modified identity operator for this case is
\begin{equation}
    \int \frac{d^np}{\sqrt{1+2\lambda p^2}}\left(\frac{\sqrt{1+2\lambda p^2}-1}{\lambda p^2}\right)^{n}|\bm{p}\rangle\langle\bm{p}|=\bm{1}.
\end{equation}
\subsubsection{Model II}
This model \cite{nouicer2005casimir} has a different choice of generic functions compared to that of (\ref{KMM generic}). This model employs the following functions:
\begin{equation}
    f(p^2)=1+\lambda p^2,\;\;\;\;\;g(p^2)=0,
\end{equation}
along with the below equations for maximally localized states:
\begin{equation}
    \kappa_i(p)=\frac{1}{p\sqrt{\lambda}}arctan\left(p\sqrt{\lambda}\right)p_i,\;\;\;\;\;\omega(p)=\frac{c}{\hbar\sqrt{\lambda}}arctan\left(p\sqrt{\lambda}\right),\;\;\;\;\;\Omega(p)=1,
\end{equation}
and the completeness property is modified as
\begin{equation}
    \int \frac{d^3p}{1+\lambda p^2}|\bm{p}\rangle\langle\bm{p}|=\bm{1}.
\end{equation}
\subsection{GUP Corrected Casimir Energy}
By incorporating the concept of minimal length and the GUP, the authors of \cite{frassino2012casimir} have derived the Hamiltonian and corrections to the Casimir Energy up to the first order in the minimal length parameter $\lambda$. The corrected form of Casimir Energy is
\begin{equation}\label{GUPenergy}
    E_i(a)=-\frac{\hbar c\pi^2}{720}\frac{S}{a^3}\Biggr[1+\Lambda_i\Biggr(\frac{\hbar^2\lambda}{a^2}\Biggr)\Biggr],
\end{equation}
where `$a$' is the distance between the plates, $S$ is the surface area of the plates and $\Lambda_i$ is a constant for $i=1,2,3$ which corresponds to different models
$$\Lambda_{1}=\pi^2\Biggl(\frac{28+3\sqrt{10}}{14}\Biggl)\;\;\;\;\;\text{(KMM)},\qquad \Lambda_{2}=4\pi^2\Biggl(\frac{3+\pi^2}{21}\Biggl)\;\;\;\;\;\text{(DGS)},\qquad \Lambda_{3}=\frac{2\pi^2}{3}\;\;\;\;\;\text{(Model II)}.$$
Thus, the expression for the attractive Casimir Force between the plates is
\begin{equation}
    F(a)=-\frac{dE}{da}=-\frac{3\hbar c\pi^2}{720}\frac{S}{a^4}\Biggr[1+\frac{5}{3}\Lambda_i\Biggr(\frac{\hbar^2\lambda}{a^2}\Biggr)\Biggr].
\end{equation}
The pressure follows
\begin{equation}
    P(a)=\frac{F}{S}=-\frac{3\hbar c\pi^2}{720}\frac{1}{a^4}\Biggr[1+\frac{5}{3}\Lambda_i\Biggr(\frac{\hbar^2\lambda}{a^2}\Biggr)\Biggr].
\end{equation}
By using the EoS relation for Casimir energy as established before, choosing $\omega=3$, we get the GUP corrected Casimir energy density
\begin{equation}\label{GUPdensity}
    \rho(a)=-\frac{\hbar c\pi^2}{720a^4}\Biggr[1+\frac{5}{3}\Lambda_i\Biggr(\frac{\hbar^2\lambda}{a^2}\Biggr)\Biggr].
\end{equation}

Note that for $\lambda=0$, we get back the usual Casimir energy density result (\ref{casimirdensity}). Throughout the paper, we'll refer to the familiar result of the uncorrected Casimir wormholes as $\lambda=0$ case, to help us draw parallels with the GUP corrected cases.

\section{The $f(T,\mathcal{T})$ formalism}\label{sec-modgrav}
We consider the Morris-Thorne metric for static and spherically symmetric wormholes, defined as
\begin{equation}\label{metric}
    ds^2=-e^{\nu(r)}dt^2+e^{\lambda(r)}dr^2+r^2d\Omega^2
\end{equation}
where $d\Omega^2=d\theta^2+sin^2(\theta)d\Phi^2$, $\nu(r)$ is the redshift function which relates to the gravitational redshift, and $e^{\lambda(r)}=\left(1-\frac{b(r)}{r}\right)^{-1}$, $b(r)$ is the shape function that determines the geometry of the wormhole. Both the redshift function and shape function must satisfy the following conditions: For the redshift, $\nu(r)$ must be non-negative and finite throughout the domain so that horizons do not exist; As $r\to\infty$, the redshift must approach zero. As for the shape function $b(r)$, the throat condition states that at the throat $r=r_0\implies b(r_0)=r_0$ and $b(r)<r$ for all $r>r_0$. Secondly, the flare-out condition $\frac{b(r)-rb'(r)}{b^2(r)}>0$ is crucial for putting constraints on the geometry around the throat. At the throat, this condition reduces to $b'(r_0)<1$. Lastly, the shape function must obey asymptotic flatness $\frac{b(r)}{r}\to\infty$ as $r\to\infty$ for a conventional Morris-Thorne wormhole.\\
The Einstein-Hilbert action integral for the $f(T, \mathcal{T})$ modified gravity is
\begin{equation}\label{action}
    S = \int e\left[\frac{1}{2k^2}f(T,\mathcal{T})+\mathcal{L}_m\right] d^4x,
\end{equation}
where $k^2$ is Einstein's gravitational constant, $\mathcal{L}_m$ represents the Matter Lagrangian Density, $e=det(e^\mu_\nu)=\sqrt{-g}$, $g$ is the determinant of the metric tensor, $\mathcal{T}$ is the trace of the Energy-Momentum tensor. The essential quantities used in this formalism are Torsion, Contorsion, and super-potential, whose components are defined respectively,
\begin{equation}\label{torsion}
    {T^\lambda}_{\mu\nu}={e_a}^\lambda(\partial_\mu{e^a}_\nu-\partial_\nu{e^a}_\mu),
\end{equation}
\begin{equation}\label{contorsion}
    {K^{\mu\nu}}_\lambda=\frac{1}{2}({T_\lambda}^{\mu\nu}+{T^{\nu\mu}}_\lambda-{T^{\mu\nu}}_\lambda),
\end{equation}
\begin{equation}\label{superpoten}
    {S_\lambda}^{\mu\nu}=\frac{1}{2}({K^{\mu\nu}}_\lambda+{\delta^\mu}_\lambda {T^{\gamma\nu}}_\gamma-{\delta^\nu}_\lambda {T^{\gamma\mu}}_\gamma).
\end{equation}
The torsion scalar can be read as
\begin{equation}\label{Tscalar}
    T={T^\lambda}_{\mu\nu}{S_\lambda}^{\mu\nu}.
\end{equation}
The standard energy-momentum tensor can be written as
\begin{equation}
    \emtensor_{\mu \nu}=-\frac{2}{\sqrt{-g}} \frac{\delta\left(\sqrt{-g} \mathcal{L}_m\right)}{\delta g^{\mu \nu}}.
\end{equation}
For our work, we consider the usual energy-momentum tensor for an anisotropic matter fluid, which is of the form
\begin{equation}\label{emtensor}
    {\emtensor_\alpha}^\beta=(\rho+p_t)u_\alpha u^\beta+p_t\delta_\alpha^\beta+(p_r-p_t)v_\alpha v^\beta,
\end{equation}
where $u_\mu=e^\frac{\nu}{2}\delta^0_\mu$ is the four-velocity and $v_\mu=e^\frac{\lambda}{2}\delta^1_\mu$ is the unit space-like vector in the radial direction. The components $\rho$ (energy density), $p_r$ (radial pressure) and $p_t$ (tangential pressure) make up the diagonal matrix ${\emtensor_\mu}^\nu=[-\rho, p_r, p_t, p_t]$.
On varying the action (\ref{action}) with respect to the metric tensor, we get the generalized field equations for this theory
\begin{equation}\label{field eq}
    [e^{-1}\partial_\epsilon(ee^\alpha_aS^{\psi\epsilon}_\alpha)+e^\alpha_aT^\epsilon_{\rho\alpha}S^{\rho\psi}_\epsilon]f_T+e^\alpha_aS^{\psi\epsilon}_\alpha(f_{TT}\partial_\epsilon T+f_{T\mathcal{T}}\partial_\epsilon\mathcal{T})+\frac{e^\psi_af}{4}-\left(\frac{e^\alpha_a\mathcal{T}^\psi_\alpha+p_te^\psi_a}{2}\right)f_\mathcal{T}=\frac{e^\alpha_a\,{\emtensor_\alpha}^\psi}{4}.
\end{equation}
Now, let us take a moment to discuss classical energy conditions derived from the Raychaudhuri equations. We will examine whether the energy conditions are fulfilled or violated by the wormhole, especially the NEC, due to the existence of exotic matter at the throat. The energy conditions for an anisotropic perfect fluid are:
\begin{itemize}
    \item Null Energy Condition (NEC): $\rho+p_r\geq0$, $\rho+p_t\geq0$
    \item Weak Energy Condition (WEC): $\rho\geq0$, $\rho+p_r\geq0$, $\rho+p_t\geq0$
    \item Dominant Energy Condition (DEC): $\rho\geq0$, $\rho-|p_r|\geq0$, $\rho-|p_t|\geq0$
    \item Strong Energy Condition (SEC): $\rho+p_r\geq0$, $\rho+p_t\geq0$, $\rho+p_r+2p_t\geq0$
\end{itemize}
all of which are satisfied by regular/normal matter due to positive energy density and pressures. However, the same cannot be said in the case of wormholes.
\section{GUP corrected Casimir Wormholes}\label{sec-Solution}
In this section, we shall discuss the wormhole solutions influenced by Casimir energy by assuming a zero tidal geometry, i.e., $\nu(r)=0$. It is evident from the above section that the choice of tetrad field will have a major impact on the equations of motion and, thus, the dynamics of gravity itself. There are two main choices for the tetrad. We shall study both of them in the following section.
\subsection{Diagonal Tetrad}
The following calculations of the field equations are based on the diagonal tetrad:
\begin{equation}
    {e^\eta}_\gamma=diag\left(e^\frac{\nu}{2}, e^\frac{\lambda}{2}, r, r\,sin\,\theta\right),
\end{equation}
whose determinant is $e=det({e^\eta}_\gamma)=e^\frac{\nu+\lambda}{2}r^2sin\,\theta$. As it has been discussed in \cite{ditta2021study}, only the linear model of $f(T, \mathcal{T})$ gravity is compatible with the diagonal tetrad, and other models of $f(T, \mathcal{T})$ theory need to follow some additional solar constraints. Thus, taking into account our choice of tetrad, we apply the following linear model,
\begin{equation}\label{model}
    f(T, \mathcal{T})=\alpha T+\beta\mathcal{T}.
\end{equation}
where $\alpha$ is a constant and $\beta$ is the coupling parameter.
The torsion scalar for the metric \eqref{metric} evaluates to
\begin{equation}\label{Tscalar}
    T=-\frac{2e^{-\lambda}}{r}\left(\nu'+\frac{1}{r}\right).
\end{equation}
Now, using Eqs. (\ref{metric}), (\ref{emtensor}), (\ref{model}) and (\ref{Tscalar}) into the equations of motion (\ref{field eq}), we can obtain the components of energy-momentum tensor
\begin{equation}\label{fullrho}
    \rho=-\frac{e^{-\lambda}}{4(\beta-1)(\beta+2)r^2}\Bigl[2\alpha\beta-4\alpha+\alpha\beta r^2\lambda'\nu'-2\alpha\beta r^2\nu''-\alpha\beta r^2(\nu')^2-\alpha\beta r\lambda'-2\alpha\beta e^\lambda-3\alpha\beta r\nu'+4\alpha r\lambda'+4\alpha e^\lambda\Bigl],
\end{equation}
\begin{equation}\label{fullpr}
    p_r=\frac{e^{-\lambda}}{4(\beta-1)(\beta+2)r^2}\Bigl[2\alpha\beta-4\alpha+\alpha\beta r^2\lambda'\nu'-2\alpha\beta r^2\nu''-\alpha\beta r^2(\nu')^2+3\alpha\beta r\lambda'-2\alpha\beta e^\lambda+\alpha\beta r\nu'+4\alpha e^\lambda-4\alpha r\nu'\Bigl],
\end{equation}
\begin{equation}\label{fullpt}
    p_t=\frac{e^{-\lambda}}{4(\beta-1)(\beta+2)r^2}\Bigl[-2\alpha\beta+\alpha r^2\lambda'\nu'-2\alpha r^2\nu''-\alpha r^2(\nu')^2+\alpha\beta r\lambda'+2\alpha\beta e^\lambda-\alpha\beta r\nu'+2\alpha r\lambda'-2\alpha r\nu'\Bigl].
\end{equation}
In this work, we consider zero tidal forces (ZTF) acting on a body near the wormhole, i.e., the redshift function $\nu(r)=0$. By this assumption and using the relation $\lambda(r)=-ln\left(1-\frac{b(r)}{r}\right)$ in (\ref{fullrho}-\ref{fullpt}), we get the following simplified field equations:
\begin{equation}\label{EFE1}
    \rho(r)=\frac{\alpha(\beta-4)r\,b'(r)+\alpha\beta\,b(r)}{4(\beta-1)(\beta+2)r^3},
\end{equation}
\begin{equation}\label{EFE2}
    p_r(r)=\frac{3\alpha\beta r\,b'(r)+\alpha(4-5\beta)\,b(r)}{4(\beta-1)(\beta+2)r^3},
\end{equation}
\begin{equation}\label{EFE3}
    p_t(r)=\frac{\alpha(\beta+2)r\,b'(r)+\alpha(\beta-2)\,b(r)}{4(\beta-1)(\beta+2)r^3}.
\end{equation}
Now, with the above expressions (\ref{EFE1}-\ref{EFE3}), we try to find the wormhole solutions under GUP-corrected Casimir energy and study the behavior of those solutions with various tools.
We replace the plate separation distance `$a$' with the radial coordinate `$r$' and solve for the shape function. This can be done, as we are dealing with Planck scale lengths, and the distance between the plates can be assumed to vary radially \cite{garattini2019casimir}.\\
Now, we equate the Casimir energy density (\ref{GUPdensity}) with the density of the modified field equation (\ref{EFE1}), and solving the differential equation, we obtain the shape function
\begin{equation}\label{sol_const}
    b(r)=\frac{{r_0}^\frac{2\beta-4}{\beta-4}}{r^\frac{\beta}{\beta-4}}-\frac{\pi^2(\beta-1)(\beta+2)}{180\alpha}\Biggl\{\frac{1}{4}\Biggl[\frac{1}{r}-\frac{{r_0}^\frac{4}{\beta-4}}{r^\frac{\beta}{\beta-4}}\Biggl]+\frac{5}{3}\cdot\frac{\Lambda_i\lambda}{12-2\beta}\Biggl[\frac{1}{r^3}-\frac{{r_0}^\frac{12-2\beta}{\beta-4}}{r^\frac{\beta}{\beta-4}}\Biggl]\Biggl\}.
\end{equation}
Here, we imposed the throat condition $b(r_0)=r_0$ to the above equation.
Now, if we define
$\gamma=-\frac{\pi^2(\beta-1)(\beta+2)}{180\alpha},$
$\sigma_i=\frac{5}{3}\cdot\frac{\Lambda_i\lambda}{12-2\beta}$,
we can make the solution look much simpler
\begin{equation}\label{simplified_b(r)}
    b(r)=\frac{{r_0}^\frac{2\beta-4}{\beta-4}}{r^\frac{\beta}{\beta-4}}+\frac{\gamma}{4}\Biggl[\frac{1}{r}-\frac{{r_0}^\frac{4}{\beta-4}}{r^\frac{\beta}{\beta-4}}\Biggl]+\gamma\sigma_i\Biggl[\frac{1}{r^3}-\frac{{r_0}^\frac{12-2\beta}{\beta-4}}{r^\frac{\beta}{\beta-4}}\Biggl].
\end{equation}
We notice that the solution consists of three terms: the geometric contribution is given by the first term, the second term is from the Casimir Effect, signifying the semiclassical quantum effects of the spacetime, and the last term is the GUP correction, which takes into account the minimal length.
Looking at (\ref{sol_const}), one could make a case for an inductive argument and extend the solution for $n-order$ GUP correction.
\begin{equation}\label{bn(r)}
    b_n(r)=\frac{{r_0}^\frac{2\beta-4}{\beta-4}}{r^\frac{\beta}{\beta-4}}-\frac{\pi^2(\beta-1)(\beta+2)}{180\alpha}\Biggl\{\sum_{j=0}^{n}\frac{D_i^{(j)}\lambda^j}{4(2j+1)-2j\beta}\Biggl[\frac{1}{r^{2j+1}}-\frac{{r_0}^\frac{4(2j+1)-2j\beta}{\beta-4}}{r^\frac{\beta}{\beta-4}}\Biggr]\Biggr\}.
\end{equation}
One may verify the above by substituting it in (\ref{EFE1}) to get the $n^{th}$ order GUP correction for Casimir energy, as briefly talked about in \cite{jusufi2020traversable}
\begin{equation}
    \rho=-\frac{\pi^2}{720r^4}\Biggr[1+D_i^{(1)}\Biggr(\frac{\sqrt{\lambda}}{r}\Biggr)^2+D_i^{(2)}\Biggr(\frac{\sqrt{\lambda}}{r}\Biggr)^4+\cdots\Biggr],
\end{equation}
where $D_i^{(j)}$ are coefficients of the correction parameter. In our case of first-order corrections $D_i^{(0)}=1$, $D_i^{(1)}=5\Lambda_i/3$. As $n$ increases, the contribution from each term decreases; hence, it is best to continue our work up to the first order.
Next up, we seek to suffice the ``asymptotic flatness" condition, $\lim_{r\to\infty}\frac{b(r)}{r}\seteq0$ which would imply that $\frac{2\beta-4}{\beta-4}>0$. This brings about case I: Both the numerator and denominator are positive: $\beta>2$\; and $\beta>4$ and case II: Both the numerator and denominator are negative: $\beta<2$\; and $\beta<4$. Also keeping in mind the presence of $\alpha$ and $12-2\beta$ in the denominators of $b(r)$,
\begin{equation}\label{asymp_ranges}
    \begin{array}{cc}
        &\alpha\in\mathbb{R}-\{0\}\\
        &\beta\in(-\infty,2)\cup(4,\infty)-\{6\}
    \end{array}
\end{equation}

\begin{figure}[H]
\centering
    \subfigure[Plot of $b(r)$]{\includegraphics[scale=0.5]{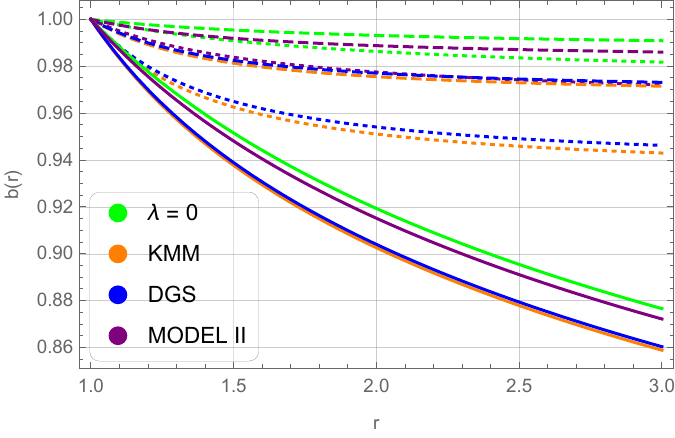}\label{b_lin}}\;\;\;
    \subfigure[Plot of $b'(r)$]{\includegraphics[scale=0.505]{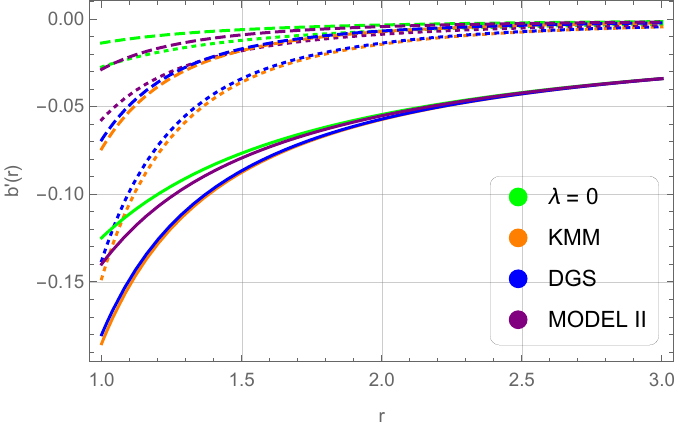}\label{b'_lin}}\;\;\;
    \subfigure[Plot of $\frac{b(r)}{r}$]{\includegraphics[scale=0.495]{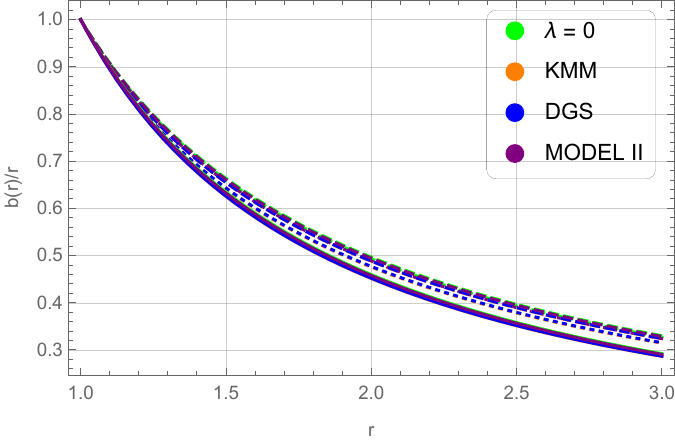}\label{b/r_lin}}
    \caption{Plot of shape function and its properties for the uncorrected case and KMM, DGS, Model II GUP corrected wormholes with $\lambda=0.1$. Dotted curves correspond to the case of GR, dashed to $\alpha =2,\;\beta=0$, and solid curves to $\alpha =2,\;\beta=-0.5$.}
    \label{fig:Shape function_lin}
\end{figure}

\begin{figure}[h]
\centering
    \subfigure[$\rho+p_r$]{\includegraphics[scale=0.395]{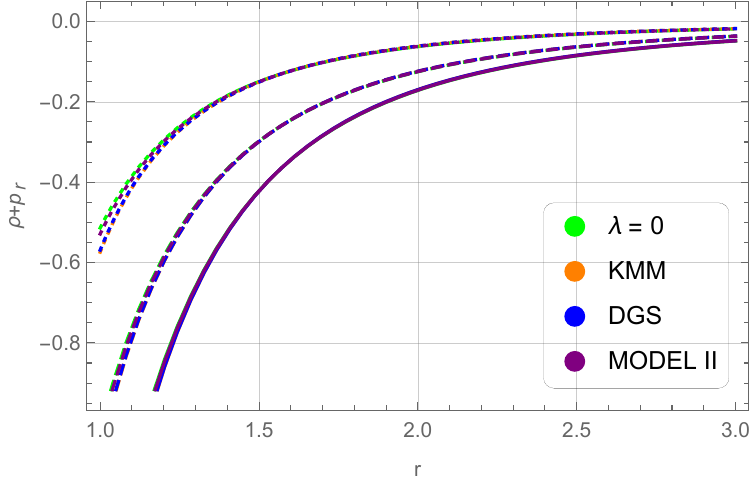}\label{}}\quad
    \subfigure[$\rho+p_t$]{\includegraphics[scale=0.39]{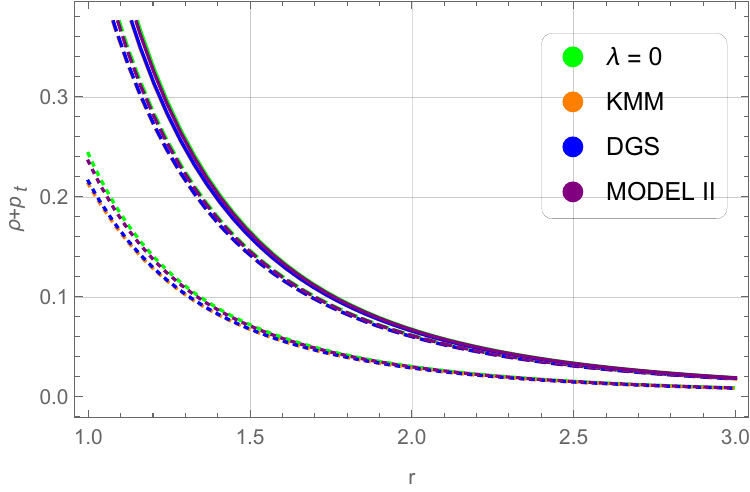}\label{}}\quad
    \subfigure[$\rho+p_r+2p_t$]{\includegraphics[scale=0.403]{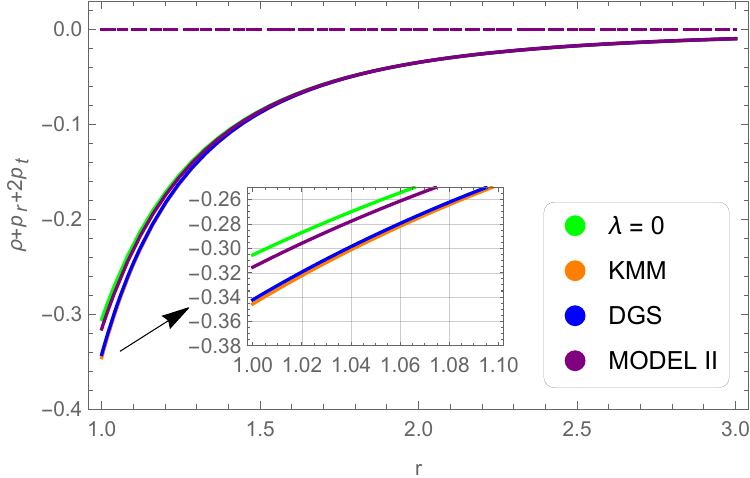}\label{}}
    \caption{Energy conditions for the uncorrected case and KMM, DGS, Model II GUP corrected wormholes with $\lambda=0.1$. Dotted curves correspond to the case of GR, dashed to $\alpha =2,\;\beta=0$, and solid curves to $\alpha =2,\;\beta=-0.5$.}
    \label{fig:EC_lin}
\end{figure}

\begin{figure}[h]
\centering
    \subfigure[Anisotropy]{\includegraphics[scale=0.395]{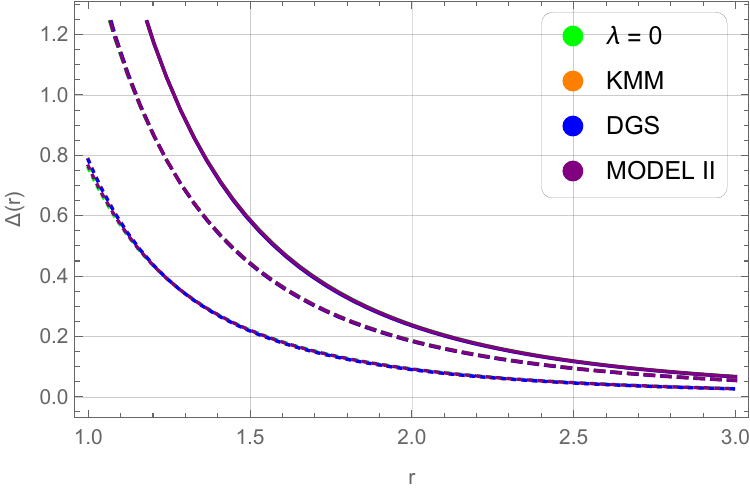}\label{}}\quad
    \subfigure[EoS parameter $\omega_r$]{\includegraphics[scale=0.39]{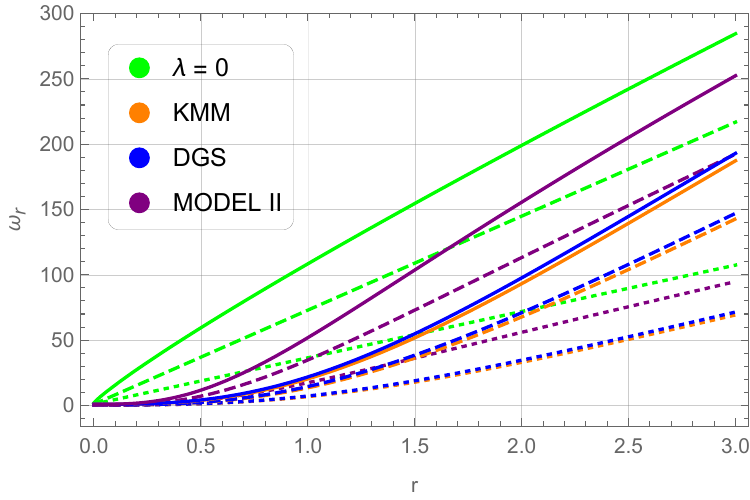}\label{}}\quad
    \subfigure[EoS parameter $\omega_t$]{\includegraphics[scale=0.403]{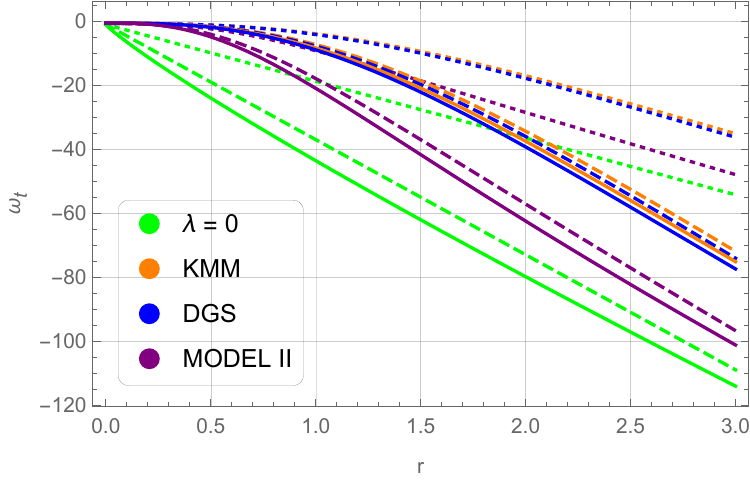}\label{}}
    \caption{Pressure anisotropy and the EoS parameters for the uncorrected case and KMM, DGS, Model II GUP corrected wormholes with $\lambda=0.1$. Dotted curves correspond to the case of GR, dashed to $\alpha =2,\;\beta=0$, and solid curves to $\alpha =2,\;\beta=-0.5$.}
    \label{fig:EC_lin2}
\end{figure}

Now, we have analyzed the properties of the shape functions shown in Fig.\ref{fig:Shape function_lin}. For our analysis, we set the throat radius at $r_0=1$ and the correction length parameter at $\lambda=0.1$ (for GUP corrected cases). Further, the functional model parameters are considered $\alpha=2$, $\beta=-0.5$, which have been assumed from the range \eqref{asymp_ranges}. 
We observed that the flare-out condition, $b'(r)$, is satisfied at the throat for each case under the asymptotic condition. The primary difference among the GUP models lies in the magnitude of the parameter $\Lambda_i$. Notably, the $\Lambda_i$ parameter is larger in the KMM approach, resulting in a slightly higher shape function value compared to the DGS model. Also, to study the deviations in the behaviors of shape functions among GR, teleparallel gravity, and teleparallel gravity with matter coupling, we plotted the graphs in the same Fig.\ref{fig:Shape function_lin}. From these comparisons, we can easily observe the influences of model parameters $\alpha$ and $\beta$ on the shape function. Furthermore, we have noticed that the GUP parameter has a significant influence beyond the wormhole throat. We also noted that as $\lambda$ increases, there is a substantial rise in the shape function within the wormhole throat. However, outside the wormhole throat, the GUP parameter decreases the shape function's value at a given radial distance.
Now, we shall study the energy conditions to study the dynamics of the wormhole solutions. We substitute the shape function (\ref{simplified_b(r)}) into the Eqs. (\ref{EFE1}-\ref{EFE3}), and obtain the expressions for NEC as follows
\begin{equation}\label{e4}
    \rho+p_r=\frac{\alpha}{(\beta+2)r^3}\Biggl(-\left(\frac{2\beta-4}{\beta-4}\right)\frac{{r_0}^\frac{2\beta-4}{\beta-4}}{r^\frac{\beta}{\beta-4}}+\gamma\Biggl\{\frac{1}{4}\Biggl[-\frac{2}{r}+\left(\frac{2\beta-4}{\beta-4}\right)\frac{{r_0}^\frac{4}{\beta-4}}{r^\frac{\beta}{\beta-4}}\Biggl]+\sigma_i\Biggl[-\frac{4}{r^3}+\left(\frac{2\beta-4}{\beta-4}\right)\frac{{r_0}^\frac{12-2\beta}{\beta-4}}{r^\frac{\beta}{\beta-4}}\Biggl]\Biggl\}\Biggr),
\end{equation}
\begin{equation}\label{e5}
    \rho+p_t=\frac{\alpha}{2(\beta+2)r^3}\Biggl(-\left(\frac{4}{\beta-4}\right)\frac{{r_0}^\frac{2\beta-4}{\beta-4}}{r^\frac{\beta}{\beta-4}}+\gamma\Biggl\{\frac{1}{\beta-4}\frac{{r_0}^\frac{4}{\beta-4}}{r^\frac{\beta}{\beta-4}}+\sigma_i\Biggl[-\frac{2}{r^3}+\left(\frac{4}{\beta-4}\right)\frac{{r_0}^\frac{12-2\beta}{\beta-4}}{r^\frac{\beta}{\beta-4}}\Biggl]\Biggl\}\Biggr),
\end{equation}
At wormhole throat $r=r_0$, the above expressions for NEC can be read as
\begin{equation}\label{e7}
    \rho+p_r\bigg\vert_{r=r_0}= -\frac{1}{540(\beta-4)(\beta+2)}\Biggl[1080\alpha(\beta-2)r_0^{-2}+3\pi^2(\beta-1)(\beta+2)r_0^{-4}\left(1+\frac{5\lambda\Lambda_i}{3r_0^2}\right)\Biggr],
\end{equation}
\begin{equation}\label{e8}
    \rho+p_t\bigg\vert_{r=r_0}= -\frac{1}{1080(\beta-4)(\beta+2)}\Biggl[2160\alpha\,r_0^{-2}+3\pi^2(\beta-1)(\beta+2)r_0^{-4}\left(1+\frac{5\lambda\Lambda_i}{3r_0^2}\right)\Biggr],
\end{equation}
where $\beta\neq -2,\,\, 4$. It is evident that the expression \eqref{e7} is a negative quantity, which confirms the violation of NEC at the throat.\\
A comprehensive graphical analysis of the energy conditions under the effect of Casimir energy and GUP models is presented in Fig.\ref{fig:EC_lin}. We fixed the model parameters from the range specified in \eqref{asymp_ranges} along with GUP parameter $\lambda=0.1$ and throat radius $r_0=1$. We checked the NEC near the throat and found that $\rho+p_r$ is violated, whereas $\rho+p_t$ is satisfied for each case.
It is interesting to mention here that the model parameters $\alpha$ and $\beta$ play a significant role in influencing all the energy conditions, especially NEC. It was observed that for positive values of $\alpha$, NEC is violated near the throat, whereas negative values give the satisfactory behavior of NEC. We also note that in violation of NEC, $f(T,\mathcal{T})$ gravity gives more contribution compared to the GR as well as $f(T)$ gravity. This is because matter coupling parameter $\beta$ plays an important role in the violation of energy conditions (for more details, see Fig.\ref{fig:EC_lin}). Further, SEC was checked, which shows negative behavior for each case.
Further, note that the quantum field theory reveals that quantum fluctuations often violate most energy conditions unrestrictedly, suggesting a significant potential influence of quantum fluctuations on the stability of wormholes. For example, one can explore the implications of the Quantum Weak Energy Condition (QWEC) constraint expressed as \cite{garattini2019casimir}
\begin{equation}
\rho(r)+p_r(r)<f(r), \quad \quad f(r)>0,
\end{equation}
where $r\in [r_0,\,\infty)$. Consequently, small deviations from energy conditions due to quantum fluctuations are possible in quantum field theory.\\
Additionally, the pressure anisotropy and EoS parameters are respectively defined as
\begin{equation}\label{e2}
    \Delta=\Delta p=p_t-p_r=\frac{\alpha(3b(r)-r\,b'(r))}{2(\beta+2)r^3},
\end{equation}
\begin{equation}\label{e3}
    \omega_r=\frac{p_r}{\rho},\;\;\;\;\;\;\;\omega_t=\frac{p_t}{\rho}.
\end{equation}
Thus, the equation for pressure anisotropy can be obtained from Eq. \eqref{e2}
\begin{equation}
    \Delta=\frac{\alpha}{2(\beta+2)r^3}\Biggl(\left(\frac{4\beta-12}{\beta-4}\right)\frac{{r_0}^\frac{2\beta-4}{\beta-4}}{r^\frac{\beta}{\beta-4}}+\gamma\Biggl\{\frac{1}{r}+\left(\frac{\beta-3}{\beta-4}\right)\frac{{r_0}^\frac{4}{\beta-4}}{r^\frac{\beta}{\beta-4}}+\sigma_i\Biggl[\frac{6}{r^3}+\left(\frac{4\beta-12}{\beta-4}\right)\frac{{r_0}^\frac{12-2\beta}{\beta-4}}{r^\frac{\beta}{\beta-4}}\Biggl]\Biggl\}\Biggr),
\end{equation}
Also, from \eqref{e3}, the EoS parameters can be read as at the throat $r=r_0$
\begin{equation}
    \omega_r(r_0)=\frac{4320\alpha(\beta-2)r_0^4+9\pi^2\beta(\beta+2)r_0^2+15\pi^2\beta(\beta+2)\lambda\Lambda_i}{\pi^2(\beta-4)(\beta+2)(3r_0^2+5\lambda\Lambda_i)},
\end{equation}
\begin{equation}
    \omega_t(r_0)=\frac{4320\alpha r_0^4+3\pi^2(\beta+2)^2r_0^2+5\pi^2(\beta+2)^2\lambda\Lambda_i}{\pi^2(\beta-4)(\beta+2)(3r_0^2+5\lambda\Lambda_i)}.
\end{equation}
Fig.\ref{fig:EC_lin2} helps us understand the behavior of the anisotropy of the spacetime around the wormhole, and  EoS parameters, which are non-constant functions of the radial coordinate $r$. We observed that the pressure anisotropy $\Delta$ is strictly positive, implying that $p_t>p_r$ guarantees the presence of exotic matter. It can be noted from its graph that $\Delta$ monotonically decreases at a rapid rate. Also, the GUP correction affects the anisotropy by increasing its value compared to that of the regular Casimir wormhole. Further, it has been noted that the modified gravity parameter significantly influences the behaviors of the EoS parameters, particularly when considered at distances away from the wormhole throat. Further observation reveals that the trajectories of the EoS parameters, denoted as $\omega_r$ and $\omega_t$, are diverging when $r \rightarrow \infty$ (see Fig.\ref{fig:EC_lin2}). This phenomenon can be attributed to the behavior outlined in Eq. \eqref{casimirdensity}, which demonstrates the Casimir force decreasing proportionally to $1/r^4$. As a result, as the distance $r$ approaches infinity, the value of $\rho$ approaches zero. This indicates that the ratios $\omega_i=\frac{p_i}{\rho}$ (where $i$ represents either $r$ or $t$) become infinite under these conditions due to $\rho\rightarrow 0$. Nevertheless, one can bypass such a result by noting that the Casimir plates we use are positioned near the wormhole throat and serve a specific role in engineering supply. Therefore, the values of $\rho$ and $p_i$ cannot be assumed for any arbitrary $r$. For this reason, our analysis and subsequent visualizations are confined to the vicinity of the throat because, at $r\rightarrow \infty$, the component of the stress-energy tensor will no longer be valid. 
\subsection{Non-Diagonal Tetrad}
\begin{figure}[H]
\centering
    \subfigure[Plot of $b(r)$]{\includegraphics[scale=0.5]{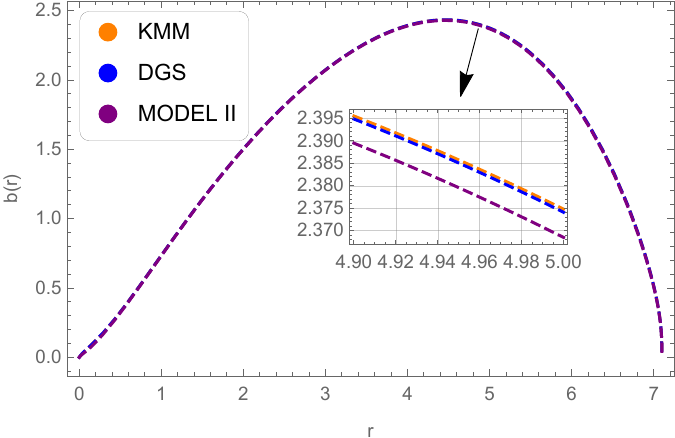}\label{b_quad}}\;
    \subfigure[Plot of $b(r)-r$]{\includegraphics[scale=0.53]{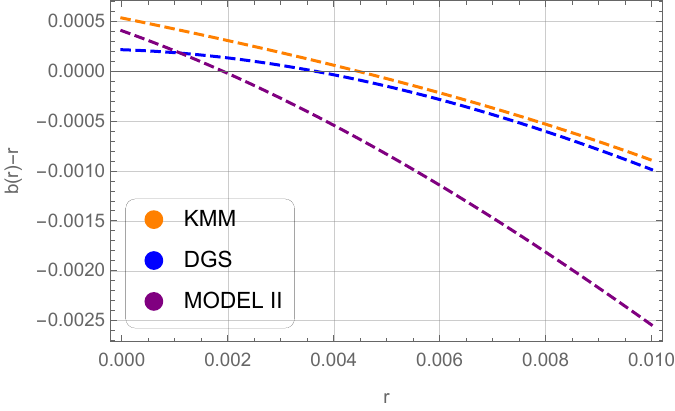}\label{b-r_quad}}\;
    \subfigure[Plot of $b'(r)$]{\includegraphics[scale=0.505]{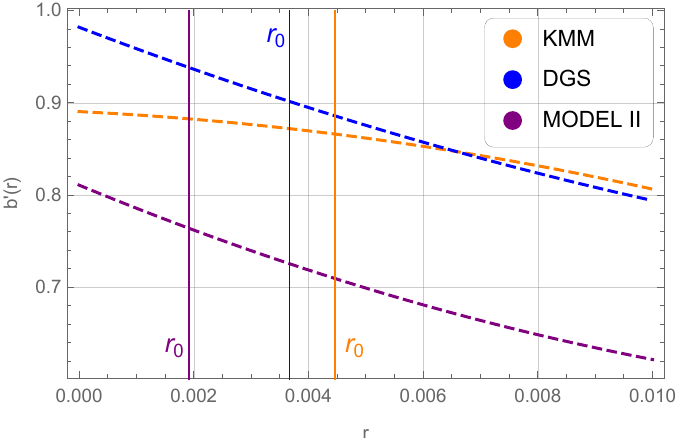}\label{b'_quad}}
    \caption{Plot of shape function and its properties for the KMM, DGS, Model II GUP corrected wormholes with $\lambda=0.1$. We have considered the model parameters as $\eta =2.5,\;\chi=-0.5$.}
    \label{fig:Shape function_quad}
\end{figure}

\begin{figure}[h]
\centering
    \subfigure[$\rho+p_r$]{\includegraphics[scale=0.5]{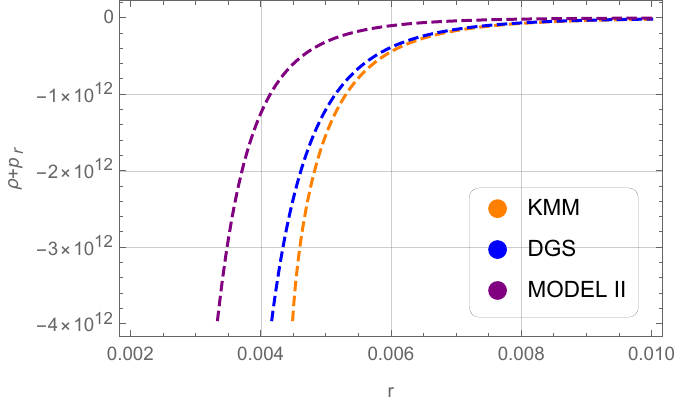}\label{}}\quad
    \subfigure[$\rho+p_t$]{\includegraphics[scale=0.5]{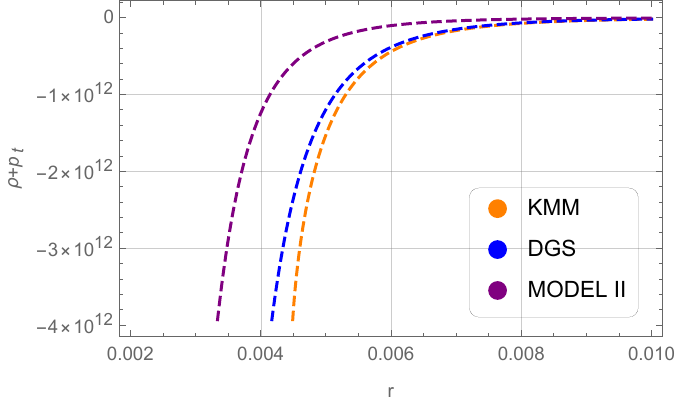}\label{}}\quad
    \subfigure[$\rho+p_r+2p_t$]{\includegraphics[scale=0.5]{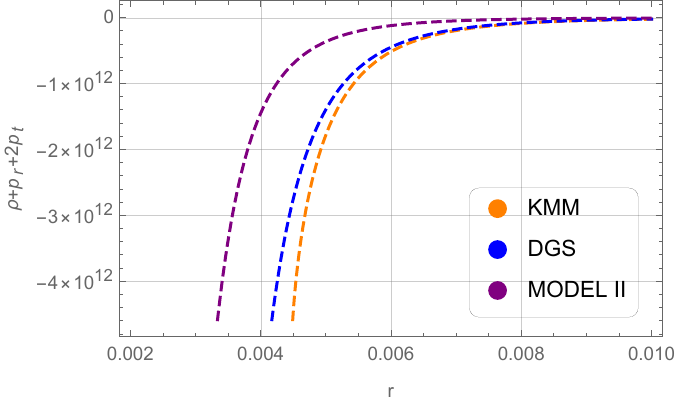}\label{}}
    \subfigure[Anisotropy]{\includegraphics[scale=0.5]{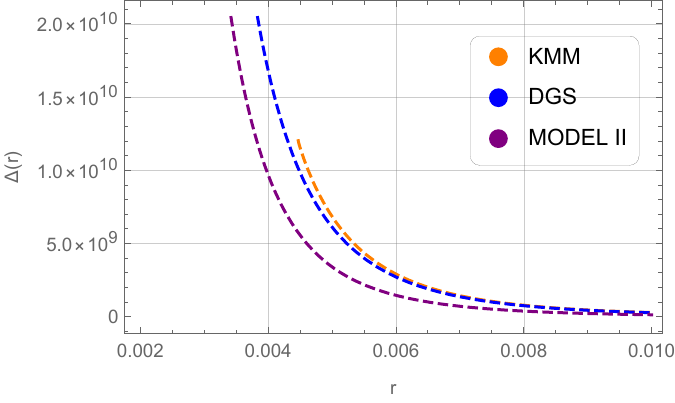}\label{}}\quad\quad
    \subfigure[EoS parameter $\omega_r$]{\includegraphics[scale=0.48]{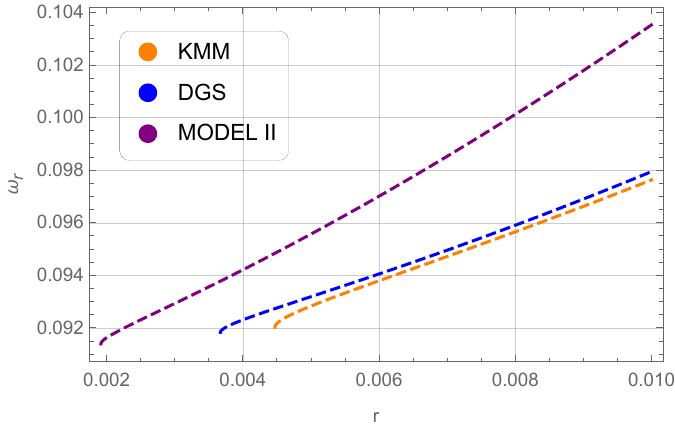}\label{}}\quad\;
    \subfigure[EoS parameter $\omega_t$]{\includegraphics[scale=0.48]{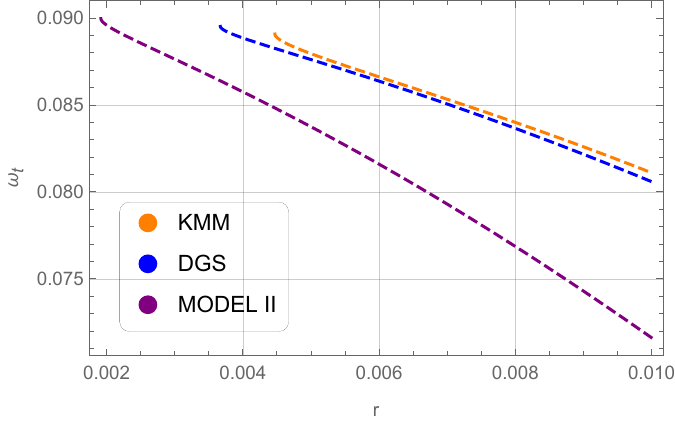}\label{}}
    \caption{NEC, SEC expressions, pressure anisotropy, and the EoS parameters for the KMM, DGS, Model II GUP corrected wormholes with $\lambda=0.1$. We have considered the model parameters as $\eta =2.5,\;\chi=-0.5$.}
    \label{fig:EC_quad}
\end{figure}

The Teleparallel Equivalent of General Relativity (TEGR) offers a consistent geometric framework for describing the dynamics of the gravitational field. The core formulation of TEGR relies on the tetrad field, which serves as a fundamental bridge connecting Dirac spinor fields with gravitational interactions. Tetrad fields also provide a comprehensive description of reference frames along line elements within a manifold. Generally, TEGR maintains invariance under the global Lorentz group $SO(3,1)$ structure without boundary conditions imposed on the tetrad fields \cite{PhysRevD.65.124001,JWMaluf_2003,PhysRevD.67.044016}. The Lagrangian density of this theory is invariant under local $SO(3,1)$ transformations up to a nontrivial, nonvanishing total divergence \cite{PhysRevD.73.124017}, and due to this, the local $SO(3,1)$ group is not a symmetry of the theory. Unlike in the Einstein–Cartan theory, which possesses local $SO(3,1)$ symmetry, gauge transformations in TEGR do not eliminate the six degrees of freedom provided by the tetrad fields, which are closely related to the metric tensor. On the other hand, the reference frame construction depends on the six components of the acceleration tensor \cite{Maluf_2007}, which plays a crucial role in defining the inertial properties of the frame. Notably, in TEGR, the tetrad framework determines both the gravitational field and the reference frame. In $f(T,\mathcal{T})$ gravity, tetrads play a significant role in formulating the field equations. In \cite{PhysRevD.86.044009}, the authors discussed that tetrads can be categorized into good and bad tetrad types. Diagonal tetrads are generally not appropriate, as they impose certain constraints on solar system observations \cite{10.1111/j.1365-2966.2012.21995.x}. In this section, we apply good tetrads within the field equations to reduce the inconsistencies associated with diagonal tetrads.\\
The non-diagonal tetrads for the spherically symmetric metric \eqref{metric} can be read as \cite{krvsvsak2016covariant}
\begin{equation}\label{proper_tetrad}
    e^\mu_\nu=
    \begin{pmatrix}
    e^{\frac{\nu}{2}} & 0 & 0 & 0\\
    0 & e^{\frac{\lambda}{2}}\text{sin}\,\theta\;\text{cos}\,\Phi & r\text{cos}\,\theta\;\text{cos}\,\Phi & -r\text{sin}\,\theta\;\text{sin}\,\Phi\\
    0 & e^{\frac{\lambda}{2}}\text{sin}\,\theta\;\text{sin}\,\Phi & r\text{cos}\,\theta\;\text{sin}\,\Phi & r\text{sin}\,\theta\;\text{cos}\,\Phi\\
    0 & e^{\frac{\lambda}{2}}\text{cos}\,\theta & -r\text{sin}\,\theta & 0\\
    \end{pmatrix}
\end{equation}
The torsion scalar $T(r)$ for the above non-diagonal tetrad \eqref{proper_tetrad} can be obtained as
\begin{equation}
    T(r)=-\frac{2e^{-\lambda}(e^{\frac{\lambda}{2}}-1)(e^\frac{\lambda}{2}-r\nu'-1)}{r^2}.
\end{equation}
The corresponding field equations of the wormhole metric \eqref{metric} for this non-diagonal tetrad can be obtained from Eqs. \eqref{emtensor}, \eqref{field eq} and \eqref{proper_tetrad} as follows:
\begin{equation}
    \rho=-\frac{e^{-\frac{\lambda}{2}}(e^{-\frac{\lambda}{2}}-1)(f_{TT}T'+f_{T\mathcal{T}}\mathcal{T}')}{r}-\frac{f_T}{2}\left(-\frac{e^{-\lambda}(1-r\lambda')}{r^2}-\frac{1}{r^2}+\frac{T}{2}\right)+\frac{f}{4}+\frac{f_\mathcal{T}}{2}(\rho+p_t),
\end{equation}
\begin{equation}
    p_r=\left(\frac{e^{-\lambda}(r\nu'+1)}{r^2}-\frac{1}{r^2}+\frac{T}{2}\right)\frac{f_T}{2}-\frac{f}{4}-\frac{f_\mathcal{T}}{2}(p_t-p_r),
\end{equation}
\begin{equation}
    p_t=\frac{e^{-\lambda}}{2}\left(-\frac{e^{\frac{\lambda}{2}}}{r}+\frac{\nu'}{2}+\frac{1}{r}\right)(f_{TT}T'+f_{T\mathcal{T}}\mathcal{T}')+\Biggl[e^{-\lambda}\left(\left(\frac{\nu'}{4}+\frac{1}{2r}\right)(\nu'-\lambda')+\frac{\nu''}{2}\right)+\frac{T}{2}\Biggr]\frac{f_T}{2}-\frac{f}{4}.
\end{equation}
For further study, we shall consider the following non-linear model
\begin{equation}\label{general_model}
    f(T,\mathcal{T})=\eta T^n+\chi\mathcal{T}+\phi,
\end{equation}
where $\eta,\chi$ and $\phi$ are model parameters along with $n\neq0$. The TEGR can be retrieved by setting $\eta=n=1$, $\chi=\phi=0$. For our study, we consider the case for $n=2$ and $\phi=0$. That is,
\begin{equation}\label{quadratic_model}
    f(T,\mathcal{T})=\eta T^2+\chi\mathcal{T}.
\end{equation}
This model was used in \cite{harko2014f} to study the cosmological implications of this theory. Incorporating this model and the assumption of $\nu(r)=0$, the individual energy-momentum tensor components yield
\begin{equation}\label{q1}
    \rho=-\mathcal{L}_1 \Biggl[(3 \chi -4)r^2 \lambda '(r)+2 \Bigl(-3 \chi +(6 \chi -8 \chi  r+10 r-8)e^{\frac{\lambda (r)}{2}} +(-3 \chi +2 \chi  r-3 r+4) e^{\lambda(r)}+9 \chi  r-11 r+4 \Bigr)\Biggr],
\end{equation}
\begin{equation}\label{q2}
    p_r=-\mathcal{L}_1\left[\chi  r^2 \lambda '(r)-2 \chi\left(e^{\frac{\lambda (r)}{2}}-1\right)^2+2 r \left(5\chi +(2-4\chi) e^{\frac{\lambda (r)}{2}}+e^{\lambda (r)}-3\right)\right],
\end{equation}
\begin{equation}\label{q3}
    p_t=\mathcal{L}_1\left[(5 \chi-6) r^2 \lambda '(r)+2 \left(\chi+2(r-\chi) e^{\frac{\lambda (r)}{2}}+(\chi-2
   \chi  r+r) e^{\lambda (r)}+(\chi-3) r\right)\right],
\end{equation}
where $\mathcal{L}_1=\frac{\eta  e^{-2 \lambda (r)} \left(e^{\frac{\lambda (r)}{2}}-1\right)^2}{(\chi -2) (\chi -1) r^5}$.
Now, we equate the above expression for density \eqref{q1} with the Casimir energy density \eqref{GUPdensity} to obtain the shape function. However, due to the resulting differential equation being non-linear with an irremovable radical sign, it cannot be solved analytically. Hence, we use the Mathematica code \textit{NDSolve} to numerically solve for the shape function $b(r)$. For this analysis, we have assumed the model parameters $\eta=2.5,\chi=-0.5$ and used the initial condition $b(7.1)=0.008$. The shape function are its relevant properties can be visualized in Fig.\ref{fig:Shape function_quad}. The shape function $b(r)$ shows increasing behavior in the neighborhood of the throat. Also, the throat radii for each correction model are found in Fig.\ref{b-r_quad}. The values of the throat radii are KMM: $r_0\approx0.0045$, DGS: $r_0\approx0.0037$, and Model II: $r_0\approx0.0019$. It is also observed that the flare-out condition is obeyed for each case. We have also noticed that the asymptotic flatness condition is met when the radius is small because of the non-linearity of the Lagrangian. It is well-established that the GUP significantly affects the semiclassical Casimir energy. Consequently, the non-linearity of the Lagrangian is unavoidable due to quantum corrections. Given these small-scale quantum corrections, it is essential to note that the asymptotic flatness condition may be satisfied far from the throat, as the GUP approximation to Casimir energy is not valid in that region. We can see a similar behavior in \cite{hassan2023gup}.
Later, we analyzed the energy conditions, especially NEC and SEC, near the wormhole throat in Fig.\ref{fig:EC_quad}. We noticed that both NEC and SEC are violated in the vicinity of the throat. Additionally, we examine the behavior of anisotropic pressure and the equation of state (EoS) parameters in Fig.\ref{fig:EC_quad}. Our observations indicate that the pressure anisotropy, $\Delta$, is strictly positive and monotonically decreasing, signifying that $p_t > p_r$ and thereby confirming the presence of exotic matter. It is also evident that the model parameters substantially affect the EoS parameter's behavior, especially at distances far from the wormhole throat. Further analysis shows that the trajectories of the EoS parameters, represented by $\omega_r$ and $\omega_t$, diverge as $r \rightarrow \infty$. A schematic note of the observations is shown in Table-\ref{tab:EC_quad}.
\begin{table}[h]
    \centering
    \begin{tabular}{|c>{\centering\arraybackslash}p{3.5cm}>{\centering\arraybackslash}p{3.5cm}>{\centering\arraybackslash}p{3.5cm}|} \hline
        \multicolumn{4}{|c|}{Overview of the energy-related stats} \\ \hline
        Expressions&  KMM&  DGS& Model II\\ \hline\hline
        $\rho+p_r$&  $negative$ at $r_0$&  $negative$ at $r_0$& $negative$ at $r_0$\\  [1ex] \hline
        $\rho+p_t$&  $negative$ at $r_0$& $negative$ at $r_0$& $negative$ at $r_0$\\  [1ex]\hline
        $\rho+p_r+2p_t$&  $negative$ at $r_0$&  $negative$ at $r_0$& $negative$ at $r_0$\\ [1ex]\hline
        $\Delta=p_t-p_r$&  $positive$ for $r\geq r_0$&  $positive$ for $r\geq r_0$& $positive$ for $r\geq r_0$\\ [1ex]\hline
        $\omega_r$&  $positive$, $increasing$&  $positive$, $increasing$& $positive$, $increasing$\\ [1ex]\hline
        $\omega_t$&  $positive$, $decreasing$&  $positive$, $decreasing$& $positive$, $decreasing$\\ [1ex]\hline
    \end{tabular}
    \caption{Behaviour of the NEC, SEC, pressure anisotropy and EoS parameters for GUP corrected Casimir wormholes with correction parameter as $\lambda=0.1$ for the non-linear model $\eta=2.5$, $\chi=-0.5$.}
    \label{tab:EC_quad}
\end{table}

\section{Quasinormal Modes}\label{sec-QNMs}
The quasinormal modes (QNMs) of compact astrophysical objects such as black holes, wormholes, etc are the characteristic oscillations of a gravitational system arising from perturbative effects. QNMs give gravity theorists and experimentalists crucial information about compact objects, allowing them to deduce possibilities of their existence along with their geometry and stability. The central part of detecting QNMs is its associated frequencies, generally expressed as a complex number, whose real parts determine the frequency of the oscillations, and the imaginary part determines the damping of these disturbances. We usually use an external field surrounding the wormhole's throat as a probe to analyze the quasinormal modes \cite{jing2005dirac,frolov2018massive,chowdhury2018quasinormal}. In this section, we shall probe the scalar, vector, axial gravitational, and Dirac field perturbations using the $3^{rd}$ order WKB approach \cite{iyer1987black}. We will neglect any possible echoes arising from quantum corrections from any region of space \cite{konoplya2022can}.

\subsection{Scalar Perturbation}
Suppose a massless scalar field $\xi$ exists around the wormhole. The dynamics of such a field are given by the Klein-Gordon equation, which, in curved spacetime, is
\begin{equation}\label{Klein-Gordon}
    \Box \xi=\frac{1}{\sqrt{-g}}\partial_\mu\left(\sqrt{-g}g^{\mu\nu}\partial_\nu\xi\right)=0.
\end{equation}
Akin to the standard methodology of solving the Schr\"odinger equation for the Hydrogen atom, we split the problem into a temporal-radial part and angular part by decomposing the scalar field as
\begin{equation}
    \xi(t,r,\theta,\phi)=\frac{1}{r}\sum_{l,m}\psi_l(t,r)Y_{lm}(\theta,\phi),
\end{equation}
where $\psi_l(t,r)$ is the radial wavefunction and $Y_{lm}(\theta,\phi)$ are the spherical harmonics indexed by $l,m$. Assuming that the time dependence of the radial wavefunction is harmonic as $\psi_l(t,r)=\hat{\psi}_l(r)e^{-i\omega t}$. Using this in \eqref{Klein-Gordon}, we  get
\begin{equation}\label{sclar_qnmDE}
    \left(\frac{d^2}{dx^2}+\omega^2-V_l(r)\right)\hat{\psi}_l(x)=0,
\end{equation}
where we define the tortoise coordinate $x$ as
\begin{equation}\label{tortoise_coord}
    dx=\pm\frac{dr}{e^{\frac{\nu(r)}{2}}\sqrt{1-\frac{b(r)}{r}}},
\end{equation}
and the resulting potential function, expressed in the radial coordinate, is
\begin{equation}
    V_l(r)=e^\nu\Biggl[\frac{l(l+1)}{r^2}-\frac{b'r-b}{2r^3}+\frac{\nu'}{2r}\left(1-\frac{b}{r}\right)\Biggr].
\end{equation}
Here, $l=1,2,3,\dots$ is called the multipole moment or the angular momentum quantum number by which the angular momentum squared is given by $L^2=l(l+1)$. The differential equation \eqref{sclar_qnmDE} is just the time-independent Schr\"odinger equation in one dimension with energy $\omega^2$ and potential $V_l(r)$. The numerical method to solve equations like these will be elaborated in the later sections.

\subsection{Electromagnetic (Vector) Perturbation}
Just as above, we begin with the equation of motion for the Electromagnetic field, which in this case is Maxwell's equation in curved spacetime,
\begin{equation}
    \left(\sqrt{-g}g^{\mu\rho}g^{\nu\sigma}F_{\rho\sigma}\right)_{,\nu}=0,
\end{equation}
where the tensor $F_{\mu\nu}=\partial_\mu A_\nu-\partial_\nu A_\mu$ represents the electromagnetic field tensor and $A_\mu$ is the electromagnetic four-potential. Given that the considered spacetime metric \eqref{metric} is static and spherically symmetric, we can decompose the vector potential $A_\mu$ using four-dimensional vector spherical harmonics \cite{zerilli1970tensor}
\begin{equation}
    A_\mu(t,r,\theta,\phi)=\int d\omega\sum_{l,m}\Biggl[a_{lm}(r)e^{-i\omega t}\begin{pmatrix}0\\0\\ \frac{1}{\text{sin}\,\theta}\partial_\phi Y_{lm}(\theta,\phi)\\ -\text{sin}\,\theta\partial_\theta Y_{lm}\end{pmatrix}_{odd}+e^{-i\omega t}\begin{pmatrix}f_{lm}(r)Y_{lm} \\u_{lm}(r)Y_{lm} \\ k_{lm}(r)\partial_\theta Y_{lm}\\ k_{lm}(r)\partial_\phi Y_{lm}\end{pmatrix}_{even}
\Biggr].
\end{equation}
As before, $l$ and $m$ denote the angular momentum and azimuthal quantum numbers, respectively. The first term on the right-hand side, i.e., the radial dependent term, has a parity $(-1)^{l+1}$, hence called the odd parity/axial term. The second term in the decomposition has the parity $(-1)^l$, thus called the even parity/polar term. Note that the decomposition has four unknown radial functions $a_{lm}(r),f_{lm}(r),u_{lm}(r)$ and $k_{lm}(r)$; nevertheless, both axial and polar terms of the decomposition yield a similar master equation:
\begin{equation}
    \frac{d^2\psi^{(1)}_{lm}}{dx^2}+\left(\omega^2-e^\nu\frac{l(l+1)}{r^2}\right)\psi^{(1)}_{lm}=0,
\end{equation}
where the associated potential of the normal mode and the master radial function $\psi^{(1)}_{lm}(r)$ are
\begin{equation}
    V_l^{(1)}(r)=e^\nu\frac{l(l+1)}{r^2},\qquad \psi^{(1)}_{lm}(r)=\begin{cases} 
      \frac{r^2}{l(l+1)}e^{-\frac{\nu(r)}{2}}\sqrt{1-\frac{b(r)}{r}}\left(-i\omega u_{lm}-\frac{df_{lm}}{dr}\right) & \text{for even parity},\\
      a_{lm} & \text{for odd parity}.
   \end{cases}
\end{equation}
It is to be noted that the function $k_{lm}$ does not appear in the final equation of motion, and the field has only two degrees of freedom, i.e., axial and polar perturbations. Such a result is consistent with the fact that a photon has two independent polarization states.

\subsection{Axial Gravitational (Tensor) Perturbation}
In this section, we will only consider axial perturbation, just as in \cite{kim2004gravitational,kim2008wormhole}. We consider the axially symmetric spacetime metric
\begin{equation}
    ds^2=-e^{2\nu_1}dt^2+e^{2\psi}(d\phi-q_1dt-q_2dr-q_3d\theta)^2+e^{2\mu_1}dr^2+e^{2\mu_2}d\theta^2.
\end{equation}
In the absence of any perturbations, the metric assumes the form
\begin{equation}
    e^{2\nu_1}=e^{2\nu},\quad e^{-2\mu_1}=1-\frac{b}{r}=\frac{\Xi}{r^2},\quad \Xi=r^2-br,\quad e^{\nu_2}=r,\quad e^{\psi}=r\,\text{sin}\,\theta,\quad q_1=q_2=q_3=0.
\end{equation}
The key feature of axial perturbations is the nonvanishing of small $q_1,q_2$ and $q_3$. In this paper, we will not consider polar perturbations with even parity, which are characterized by linearly perturbed $\delta\nu_1,\delta\psi,\delta\mu_1,\delta\mu_2$. We follow a similar pattern as above, the equations of motion in this case are given by Einstein's equations (if the time dependence of $q_i$'s is $e^{i\omega t}$), which yield
\begin{align}
    \frac{e^\nu}{\sqrt{\Xi}}\frac{1}{r^3\text{sin}^3\,\theta}\frac{\partial Q}{\partial\theta}=&-i\omega q_{1,2}-\omega^2q_2,\\
    \frac{e^\nu\sqrt{\Xi}}{r^3\text{sin}^3\,\theta}\frac{\partial Q}{\partial r}=&+i\omega q_{1,3}+\omega^2q_3.
\end{align}
Let $Q(r,\theta)=Q(r)C^{-3/2}_{l+2}(\theta)$, where the Gegenbauer function $C^m_n(\theta)$ satisfies the following differential equation
\begin{equation}
    \Biggl[\frac{d}{d\theta}\text{sin}^{2m}\,\theta\frac{d}{d\theta}+n(n+2m)\text{sin}^{2m}\,\theta\Biggr]C^m_n(\theta)=0.
\end{equation}
If $Q(r)=rZ(r)$ and using the tortoise coordinate for the derivatives, the resulting wave equation is
\begin{equation}
    \left(\frac{d^2}{dx^2}+\omega^2-V^{(-)}\right)Z=0
\end{equation}
The potential in this equation is,
\begin{equation}
    V^{(-)}(r)=e^\nu\Biggl[\frac{l(l+1)}{r^2}+\frac{b'r-5b}{2r^3}-\frac{\nu'}{2r}\left(1-\frac{b}{r}\right)\Biggr].
\end{equation}
One may notice the general pattern here. The general form of the potential function for the perturbation of a field associated with a boson of spin $`s'$ is
\begin{equation}
    V(r)=e^\nu\frac{l(l+1)}{r^2}+e^\nu(1-s)\Biggl[\frac{\nu'}{2r}\left(1-\frac{b}{r}\right)+\frac{(1+2s)b-b'r}{2r^3}\Biggr].
\end{equation}
For each value of $s$, we acquire the potential function associated with the wave equation governing the QNMs for that field. e.g., for the scalar field, the intrinsic spin is $s=0$, which is found to be the case for particles such as Higgs Boson or Pions. For the vector field, $s=1$, which relates to Electromagnetism, whose associated boson is the photon. For the tensor field, the hypothetical graviton is expected to have the spin $s=2$, reducing to the case of axial gravitational perturbation and so on. Moving forward, we examine the Fermionic or Dirac field perturbation.

\subsection{Dirac Perturbation}
The general equation for the dynamics of a Dirac field $\Upsilon$ with mass $m$ is is given by \cite{brill1957interaction,cho2003dirac}
\begin{equation}
    [\gamma^a{e_a}^\mu(\partial_\mu+\Gamma_\mu)+m]\Upsilon=0.
\end{equation}
Here $\Gamma_\mu=\frac{1}{8}[\gamma^a,\gamma^b]e_a^\nu e_{b\nu;\mu}$ is the spin connection, $\gamma^a$ are the Dirac matrices, and ${e_a}^\mu$ is the inverse tetrad. On using the diagonal tetrad and assuming the field is massless ($m=0$), we get
\begin{equation}
    \partial_x^2\Upsilon_l(x)+\omega^2\Upsilon_l(x)=V_{d\pm}(r)\Upsilon_l(x).
\end{equation}
For the Dirac field, the two isospectral potentials are
\begin{equation}
    V_{d\pm}(r)=\frac{k}{r}\left(\frac{ke^\nu}{r}\mp \frac{e^\nu\sqrt{1-\frac{b}{r}}}{r}\pm e^{\frac{\nu}{2}}\sqrt{1-\frac{b}{r}}\frac{de^{\frac{\nu}{2}}}{dr}\right),
\end{equation}
where $k=1,2,3\hdots$ are called multipole numbers with $k=l+1/2$. These two potentials can be transformed from one to another using the Darboux transformation:
\begin{equation}
    \Upsilon_{+}=A\left(W+\frac{d}{dx}\right)\Upsilon_{-},\qquad W=\sqrt{e^{\nu/2}\sqrt{1-\frac{b}{r}}}.
\end{equation}
Here, $A$ is a constant. As $+$ and $-$ wave equations are isospectral, we can consider only one of the two effective potentials for the Dirac case.

\subsection{WKB method}
\begin{figure}[h]
\centering
    \subfigure[$V_l(x)$ for linear model, $l=5$]{\includegraphics[scale=0.5]{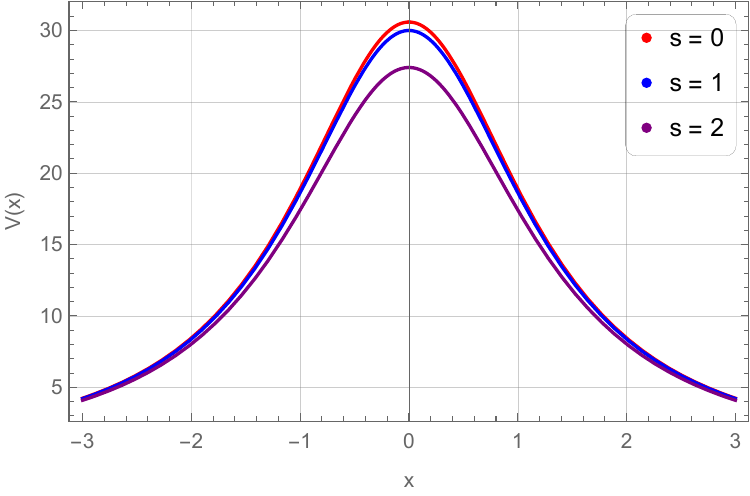}\label{}}\quad
    \subfigure[$V_l(x)$ for non-linear model, $l=5$]{\includegraphics[scale=0.55]{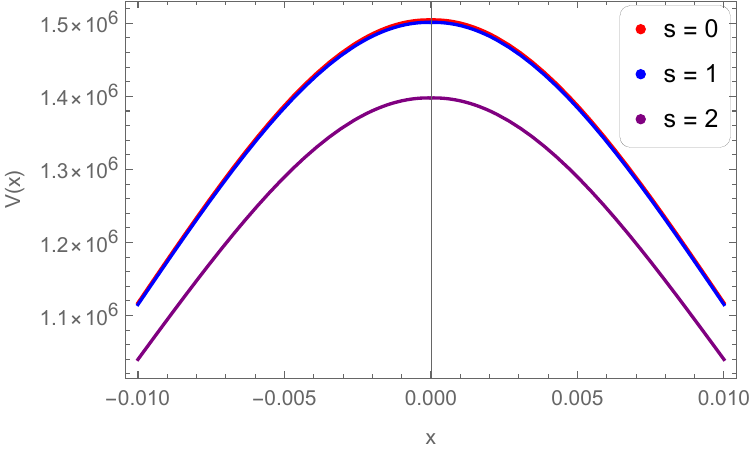}\label{}}\quad
    \subfigure[$V_d(x)$ for linear model]{\includegraphics[scale=0.5]{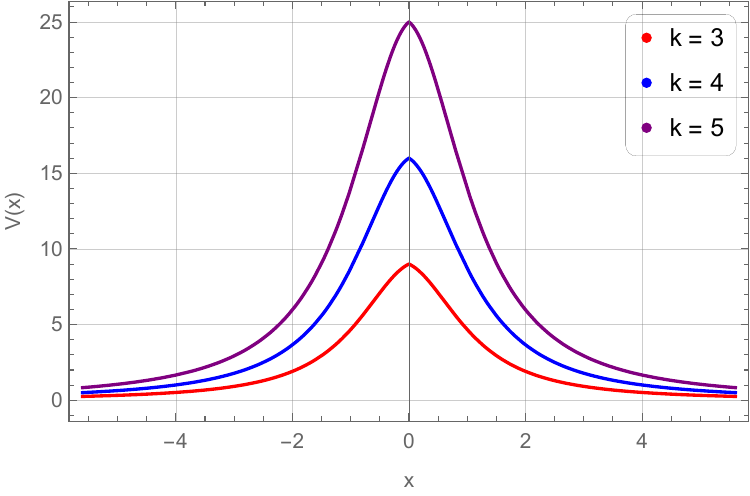}\label{}}
    \subfigure[$V_d(x)$ for non-linear model]{\includegraphics[scale=0.545]{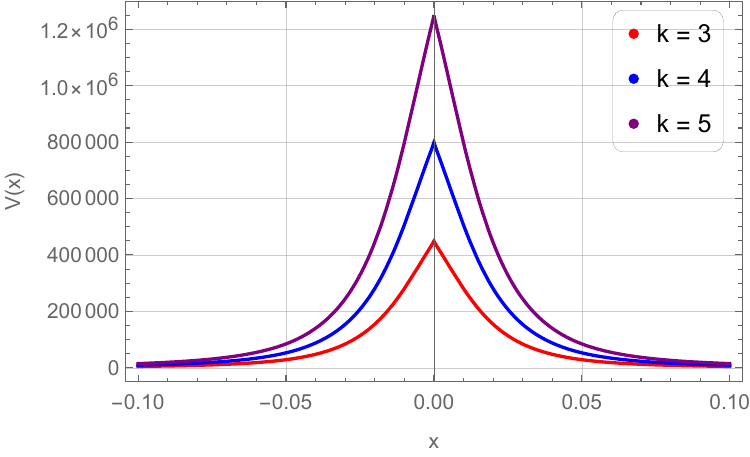}\label{}}
    \caption{Potential function for the quasinormal modes as a function of tortoise coordinate $x$. We have considered the model parameters as $\alpha=2,\,\eta =2.5,\,\beta=\chi=-0.5$.}
    \label{fig:QNM_poten}
\end{figure}

The Wentzel-Kramers-Brillouin (WKB) approximation is a well-known mathematical technique to solve linear differential equations in the context of semiclassical physics. The use of the WKB method for the calculation of quasinormal modes in the first order was first suggested by Schutz and Will \cite{schutz1985black}. Higher-order methods were later forged following the effectiveness of this method \cite{iyer1987black2,konoplya2003quasinormal}. For the scope of our paper, we will calculate the quasinormal frequencies up to third-order WKB expansion \cite{iyer1987black}. As discussed above, the modus operandi for calculating the QNM frequency is to first reduce the equations of motion of a given field to a one-dimensional time-independent Schr\"odinger's wave-like equation using the tortoise coordinate, i.e.,
\begin{equation}\label{red_wave_eq}
    \frac{d^2\Psi}{dx^2}+Q(\omega,x)\Psi=0.
\end{equation}
The function $Q(\omega,x)$ commonly, though not exclusively, appears in the form $Q(\omega,x)=\omega^2-V(x)$, where $V$ is the potential function. The (square of) QNM frequencies $\omega^2$ in the WKB approximation, carried to third order beyond
the eikonal approximation is, 
\begin{equation}\label{WKB_approx}
    \omega^2=[V_0+(-2V_0'')^{1/2}\widetilde{\Lambda}]-i\left(n+\frac{1}{2}\right)(-2V_0'')^{1/2}(1+\widetilde{\Omega}),
\end{equation}
where the terms $\widetilde{\Lambda}$ and $\widetilde{\Omega}$ are defined as
\begin{equation}
    \widetilde{\Lambda}(n)=\frac{1}{(-2V_0'')^{1/2}}\Biggl\{\frac{1}{8}\Biggl[\frac{V_0^{(4)}}{V_0''}\Biggr]\left(\frac{1}{4}+\alpha^2\right)-\frac{1}{288}\Biggl[\frac{V_0'''}{V_0''}\Biggr]^2\left(7+60\alpha^2\right)\Biggr\},
\end{equation}
\begin{multline}
    \widetilde{\Omega}(n)=\frac{1}{(-2V_0'')}\Biggl\{\frac{5}{6912}\Biggl[\frac{V_0'''}{V_0''}\Biggr]^4\left(77+188\alpha^2\right)-\frac{1}{384}\Biggl[\frac{V_0'''^2 V_0^{(4)}}{V_0''^3}\Biggr]\left(51+100\alpha^2\right)+\frac{1}{2304}\Biggl[\frac{V_0^{(4)}}{V_0''}\Biggr]^2\left(67+68\alpha^2\right)\\+\frac{1}{288}\Biggl[\frac{V_0''' V_0^{(5)}}{V_0''^2}\Biggr]\left(19+28\alpha^2\right)-\frac{1}{288}\Biggl[\frac{V_0^{(6)}}{V_0''}\Biggr]\left(5+4\alpha^2\right)\Biggr\}.
\end{multline}
In the above expressions, the primes and the superscript $(i)$ denote the order of differentiation with respect to $x$. The naughts in the subscript signify that the variables are evaluated at $x=x_0$, i.e. the location of the peak of the potential $V'(x_0)=0$. Here $\alpha$ is given by
\begin{equation}
    \alpha=n+\frac{1}{2},\qquad n=\begin{cases} 
      0,1,2,\hdots, & Re(\omega)>0,\\
      -1,-2,-3,\hdots, & Re(\omega)<0 
   \end{cases}
\end{equation}
where $n$ is called the overtone number. In the case of tideless wormholes and many uncommon cases, the potential's peak is reached at the wormhole throat $r_0$. Mathematically, this problem is very similar to studying one-dimensional quantum mechanical scattering near the peak of the potential barrier and determining the scattering resonances. The potential functions for the quasinormal modes have been plotted in Fig.\ref{fig:QNM_poten}. The peak is reached at the wormhole throat, corresponding to $x=0$. The regions $x>0$ and $x<0$ correspond to distances (to the throat) measured from the two universes connected by the throat. The potential falls off as the distance from the throat increases, alluding to the fact that at large distances, the solutions to the wave equation \eqref{red_wave_eq} are asymptotically plane waves. Thus the following boundary condition is often used for solving the wave equation:
\begin{equation}
    \Psi(x)\sim e^{\pm i\omega x},\qquad x\to\pm\infty
\end{equation}

\begin{figure}[h]
\centering
    \subfigure[Linear model]{\includegraphics[scale=0.5]{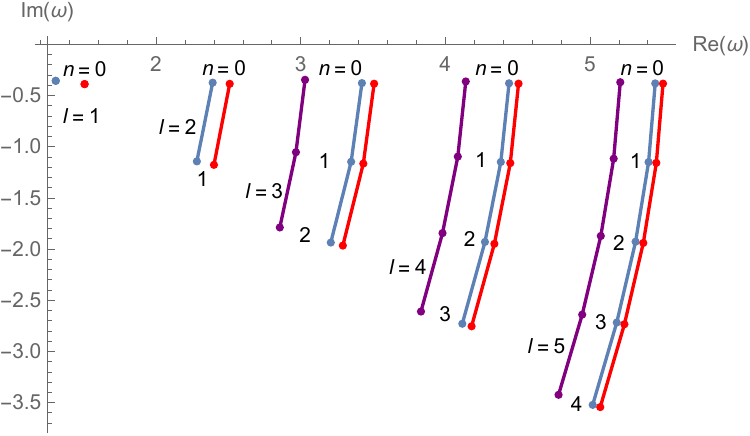}\label{}}\quad
    \subfigure[Non-linear model]{\includegraphics[scale=0.5]{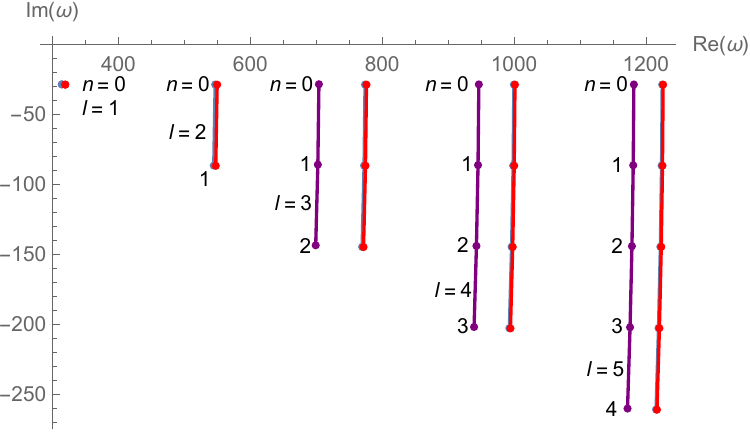}\label{}}
    \caption{Plot of quasinormal frequencies. Red plots correspond to scalar ($s=0$), blue to electromagnetic ($s=1$), and purple to gravitational ($s=2$) perturbations. KMM correction is assumed.}
    \label{fig:QNM_boson_freq}
\end{figure}
\begin{figure}[h]
\centering
    \subfigure[Linear model]{\includegraphics[scale=0.5]{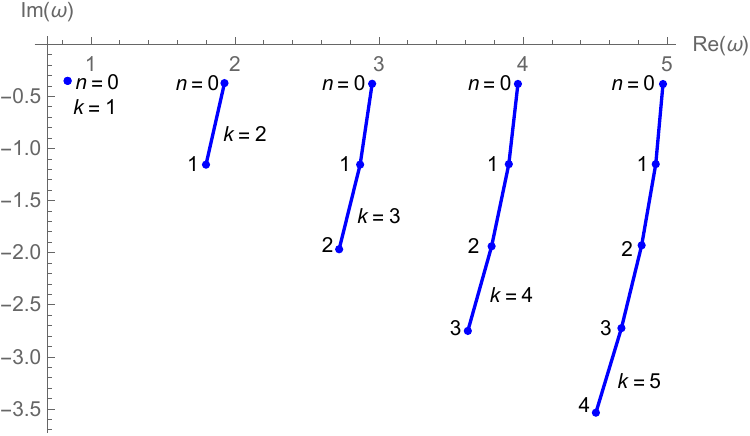}\label{}}
    \subfigure[Non-linear model]{\includegraphics[scale=0.5]{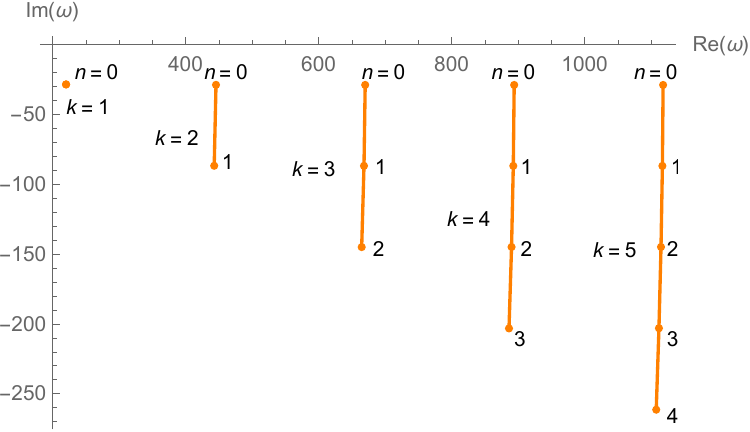}\label{}}
    \caption{Plot of quasinormal frequencies for Dirac perturbations (KMM correction is assumed).}
    \label{fig:QNM_dirac_freq}
\end{figure}

\indent When a compact astrophysical object is perturbed, it attempts to return to its original state by losing energy through normal mode oscillations. For a stable system, these oscillations are damped and eventually die out exponentially, which is the case for $Im(\omega)<0$. On the other hand, if the imaginary part of the QNM frequency is positive, it indicates that the system is unstable and the perturbations grow exponentially over time. This will result in the pinching off of the throat. Using the above WKB approximation up to the $3^{rd}$ order, we have calculated these frequencies for different quantum states, and present it in a tabulated format in Tables-\ref{tab:QNM_boson_lin},\ref{tab:QNM_boson_quad} and \ref{tab:QNM_dirac}. Subsequently, we plot these frequencies in the complex plane for better visualization (see Figs.\ref{fig:QNM_boson_freq},\ref{fig:QNM_dirac_freq}). Both the wormhole geometries, for the linear and non-linear models, are proved to be stable under any form of quantum or semiclassical perturbations as their QNM frequencies are negative. We also observe that for a fixed multipole moment $l$ (or $k$ in the Dirac case), the frequency of oscillations decreases, and the damping rate increases for increasing overtones. This signifies that, for any gravitational wave detectors like the upcoming LISA, most of the oscillations are felt from the fundamental tone with the subsequent overtones contributing less.

\begin{table}[h]
    \centering
    \begin{tabular}{|cc|c||c||c|} \hline 
  \multicolumn{2}{|c|}{Multipole moment and Overtones}& \multicolumn{3}{|c|}{Quasinormal Frequencies $\omega=Re(\omega)+Im(\omega)i$}\\ \hline  \hline
         $\qquad\qquad l$&$n$&$s=0$&$s=1$ &$s=2$\\ \hline  
           $\qquad\qquad1$&  $0$&  $1.50649-0.390084i$& $1.30702-0.358103i$&$-$\\ \hline  
           $\qquad\qquad2$&  $0$&  $2.50978-0.387291i$& $2.39192-0.375756i$&$-$\\
           &  $1$&  $2.40095-1.17894i$& $2.28282-1.14495i$&$-$\\ \hline  
           $\qquad\qquad3$&  $0$&  $3.50844-0.386307i$&   $3.42413-0.3803i$&$3.03083-0.348104i$\\ 
           &  $1$&  $3.43372-1.16668i$&   $3.34965-1.14883i$&$2.96792-1.05558i$\\
           &  $2$&  $3.29219-1.96736i$&   $3.20889-1.93803i$&$2.85635-1.79036i$\\ \hline  
           $\qquad\qquad4$&  $0$&  $4.50706-0.385812i$&   $4.4414-0.382132i$&$4.14228-0.364382i$\\
           &  $1$&  $4.45012-1.16189i$&   $4.38463-1.15092i$&$4.08716-1.09817i$\\  
           &  $2$&  $4.33989-1.95027i$&   $4.27484-1.93218i$&$3.98142-1.84564i$\\  
  & $3$& $4.18247-2.75601i$&  $4.11826-2.73097i$&$3.83234-2.61188i$\\ \hline 
  $\qquad\qquad5$& $0$& $5.50599-0.385533i$&  $5.45222-0.383052i$&$5.21048-0.371796i$\\
  & $1$& $5.45992-1.1595i$&  $5.40626-1.15209i$&$5.16486-1.11847i$\\ 
  & $2$& $5.36976-1.9417i$&  $5.31636-1.92945i$&$5.07584-1.87384i$\\ 
  & $3$& $5.23905-2.73618i$&  $5.18612-2.71919i$&$4.94727-2.64208i$\\ 
  & $4$& $5.07202-3.54521i$&  $5.01983-3.52357i$&$4.78373-3.42536i$\\ \hline 
    \end{tabular}
    \caption{Table of complex quasinormal frequencies for scalar, vector, and tensor perturbations for the linear model.}
    \label{tab:QNM_boson_lin}
\end{table}

\begin{table}[h]
    \centering
    \begin{tabular}{|cc|c||c||c|} \hline 
  \multicolumn{2}{|c|}{Multipole moment and Overtones}& \multicolumn{3}{|c|}{Quasinormal Frequencies $\omega=Re(\omega)+Im(\omega)i$}\\ \hline  \hline
         $\qquad\qquad l$&$n$&$s=0$&$s=1$ &$s=2$\\ \hline  
           $\qquad\qquad1$&  $0$&  $319.387-28.9483i$& $314.123-28.7711i$&$-$\\ \hline  
           $\qquad\qquad2$&  $0$&  $549.742-28.9484i$& $546.691-28.889i$&$-$\\
           &  $1$&  $547.634-86.9617i$& $544.579-86.7832i$&$-$\\ \hline  
           $\qquad\qquad3$&  $0$&  $776.214-28.9483i$&   $774.054-28.9185i$&$704.079-28.6804i$\\ 
           &  $1$&  $774.727-86.9025i$&   $772.566-86.8132i$&$702.45-86.1093i$\\
           &  $2$&  $771.779-145.026i$&   $769.616-144.877i$&$699.223-143.737i$\\ \hline  
           $\qquad\qquad4$&  $0$&  $1001.44-28.9483i$&   $999.771-28.9304i$&$946.628-28.8i$\\
           &  $1$&  $1000.29-86.8792i$&   $998.621-86.8255i$&$945.415-86.438i$\\  
           &  $2$&  $998.006-144.912i$&   $996.331-144.823i$&$943.003-144.188i$\\  
  & $3$& $994.602-203.11i$&  $992.926-202.985i$&$939.418-202.119i$\\ \hline 
  $\qquad\qquad5$& $0$& $1226.12-28.9482i$&  $1224.75-28.9363i$&$1181.77-28.8531i$\\
  & $1$& $1225.18-86.8676i$&  $1223.81-86.8318i$&$1180.8-86.5837i$\\ 
  & $2$& $1223.31-144.855i$&  $1221.94-144.795i$&$1178.86-144.387i$\\ 
  & $3$& $1220.52-202.954i$&  $1219.15-202.87i$&$1175.97-202.308i$\\ 
  & $4$& $1216.83-261.205i$&  $1215.46-261.097i$&$1172.16-260.391i$\\ \hline 
    \end{tabular}
    \caption{Table of complex quasinormal frequencies for scalar, vector, and tensor perturbations for the non-linear model.}
    \label{tab:QNM_boson_quad}
\end{table}

It is also noteworthy that for the wormhole geometry in the non-linear case \eqref{quadratic_model}, the frequencies of each overtone and its corresponding damping rate are extremely high in comparison with the results for the solution \eqref{sol_const}. This can be attributed to the non-linearity of the Lagrangian and the quantum corrections. Since the general structure of black holes and wormholes are similar, it is worth reading up on the QNMs for black holes in the following articles \cite{konoplya2003quasinormal,kanti2006quasinormal,konoplya2011quasinormal}. The quasinormal frequencies are studied for the Ellis–Bronnikov wormholes in \cite{kim2008wormhole,dutta2020revisiting}. In \cite{biswas2022echoes}, the authors have found that the braneworld wormholes can sustain without exotic matter, unlike in regular four-dimensional spacetime. The key study in this article is the analysis of QNMs, in which they have made a vis-à-vis comparison to normal modes of black holes. It was found that the QNM frequencies are extremely sensitive to the braneworld parameter as the damping for higher dimensional wormholes is extremely low, which could make or break the throat. Recently, an article on tideless wormholes surrounded by clouds of strings in the context of $f(R)$ gravity \cite{gogoi2023tideless} has also made observations on the QNM frequencies for scalar and Dirac perturbations by varying model parameters and using $5^{th}$ order WKB expansion.
\begin{table}[h]
    \centering
    \begin{tabular}{|cc||c||c|} \hline 
         \multicolumn{2}{|c|}{Multipole moment and Overtones}&  \multicolumn{2}{|c|}{Quasinormal Frequencies $\omega=Re(\omega)+Im(\omega)i$}\\ \hline\hline
         $\qquad\qquad k$&  $n$&  Linear Model& Non-linear Model\\ \hline 
         $\qquad\qquad1$&  $0$&  $0.839607-0.352145i$& $220.392-28.8471i$\\ \hline 
         $\qquad\qquad2$&  $0$&  $1.92833-0.373629i$& $445.789-29.1334i$\\
         &  $1$&  $1.80013-1.15647i$& $443.091-87.0107i$\\ \hline 
         $\qquad\qquad3$&  $0$&  $2.95328-0.380035i$& $670.052-29.1867i$\\
         &  $1$&  $2.86992-1.1559i$& $668.263-87.0732i$\\
         &  $2$&  $2.72428-1.96865i$& $664.792-145.176i$\\ \hline 
         $\qquad\qquad4$&  $0$&  $3.96507-0.381607i$& $894.038-29.2053i$\\
         &  $1$&  $3.90191-1.15173i$& $892.699-87.0959i$\\
 & $2$& $3.7825-1.9402i$&$890.089-145.109i$\\
 & $3$& $3.61744-2.75249i$&$886.24-203.323i$\\ \hline 
 $\qquad\qquad5$& $0$& $4.97206-0.382608i$&$1117.92-29.2139i$\\ 
 & $1$& $4.92168-1.15157i$&$1116.85-87.1065i$\\
 & $2$& $4.82384-1.93102i$&$1114.75-145.078i$\\
 & $3$& $4.68349-2.72561i$&$1111.66-203.18i$\\
 & $4$& $4.50627-3.53736i$&$1107.59-261.459i$\\ \hline
    \end{tabular}
    \caption{Table of complex quasinormal frequencies for Dirac perturbations for both the linear and non-linear models.}
    \label{tab:QNM_dirac}
\end{table}
One may extend the above findings by considering higher-order WKB approximations and time-domain analysis, which we have not included in this paper.

\section{Physical Properties of GUP corrected Casimir Wormholes}\label{sec-physicalprops}
In this section, we study a few essential physical properties of the obtained wormhole solutions, such as active mass function, total gravitational energy, volume integral quantifier (VIQ), stability analysis through the Tolman–Oppenheimer–Volkoff (TOV) equation, and the embedding diagrams of the wormholes.
\subsection{Active Mass Function}
\begin{figure}[H]
    \centering
    \includegraphics[scale=0.5]{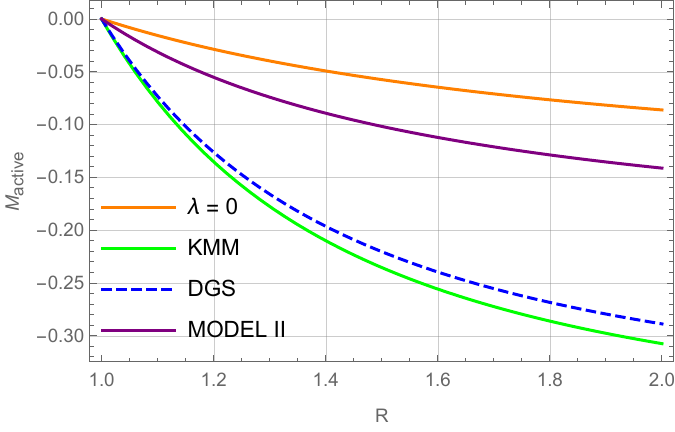}
    \caption{Active gravitational mass for the uncorrected case and the KMM, DGS, Model II GUP corrected wormholes with $\lambda=0.1$.}
    \label{fig:gravmass}
\end{figure}

The active gravitational mass quantifies the strength of gravitational flux of the astrophysical object, in our context a wormhole. The active mass of a wormhole within the region from the throat $r_0$ to an outer radial distance $R$ is
\begin{align}
    M_{active}=&\notag\int_{r_0}^R4\pi\rho r^2\,dr.
\end{align}
The expression of $M_{active}$ for GUP-corrected Casimir wormhole can be read as
\begin{equation}
 M_{active}=\frac{\pi^3}{180}\Biggl[\left(\frac{1}{R}-\frac{1}{r_0}\right)+\frac{5}{9}\Lambda_i\lambda\left(\frac{1}{R^3}-\frac{1}{r_0^3}\right)\Biggr].
\end{equation}
We have depicted the behavior of $M_{active}$ for the usual Casimir wormhole as well as the GUP-corrected Casimir wormholes in Fig.\ref{fig:gravmass}. 
One can notice from Fig.\ref{fig:gravmass} active gravitational mass $M_{active}$ decreases as we increase `$r$'. The negative behavior of $M_{active}$ is expected due to the presence of the Casimir energy, and this behavior indicates the presence of exotic matter at the throat. This type of matter is known to violate energy conditions, especially NEC and its effects we already observed in the previous section.
\subsection{Total Gravitational Energy}
\begin{figure}[H]
    \centering
    \subfigure[Linear case with $\alpha =2,\;\beta=-0.5$]{\includegraphics[scale=0.7]{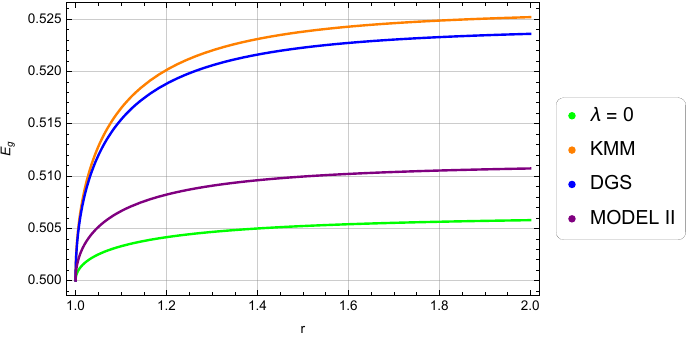}\label{}}
    \subfigure[Non-linear case with $\eta =2.5,\;\chi=-0.5$]{\includegraphics[scale=0.74]{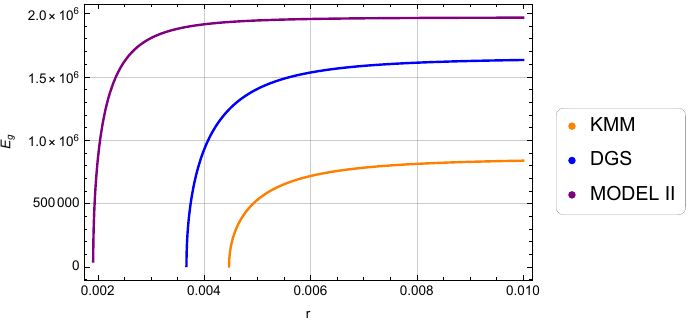}\label{}}
    \caption{Total gravitational energy of the wormhole geometry.}
    \label{fig:grav_energy}
\end{figure}

The total gravitational energy, in essence, gives us information about the gravitational effects characterized by the wormhole's geometry and mass distribution.
We investigate the total gravitational energy of the wormhole, following the works of Lyndell-Bell et al. \cite{lynden2007energy}, Katz et al. \cite{katz2006gravitational}, and Nandi et al. \cite{nandi2009gravitational}, which define it as
\begin{equation}\label{grav_energy}
    E_g=Mc^2-E_M,
\end{equation}
where
\begin{equation}
    Mc^2=\frac{1}{2}\int_{r_0}^r\rho r^2dr+\frac{r_0}{2},\;\;\;\;E_M=\frac{1}{2}\int_{r_0}^r\sqrt{g_{rr}}\rho r^2dr.
\end{equation}
The first term is the total energy, while the second is the total mechanical energy, which encapsulates kinetic energy, rest energy, internal energy, etc. All in all, (\ref{grav_energy}) modifies as
\begin{equation}\label{gravitational_energy}
    E_g=\frac{1}{2}\int_{r_0}^r(1-\sqrt{g_{rr}})\rho r^2dr+\frac{r_0}{2},
\end{equation}
with $g_{rr}=\left(1-\frac{b(r)}{r}\right)^{-1}$.\\
Since it is not possible to evaluate this integral analytically, we solve it numerically for the uncorrected and GUP-corrected cases. 
Fig.\ref{fig:grav_energy} shows that $E_g>0$, which means that there is repulsion around the throat. Such behavior is anticipated due to the presence of exotic matter, essential for a physically valid traversable wormhole.

\begin{table}[H]
    \centering
    \begin{tabular}{ |p{1.5cm}| p{1.5cm} p{1.5cm} p{1.5cm} p{1.5cm}| }
     \hline
     \multicolumn{5}{|c|}{The values of $E_g$ for different $r$} \\
     \hline
     $r$ & $\lambda=0$ & KMM & DGS & Model II\\
     \hline
     1.1 & 0.503278 & 0.516454 & 0.515370 & 0.506631\\
     1.2 & 0.504144 & 0.520118 & 0.518805 & 0.508211\\
     1.3 & 0.504644 & 0.521975 & 0.520551 & 0.509057\\
     1.4 & 0.504974 & 0.523071 & 0.521583 & 0.509582\\
     1.5 & 0.505207 & 0.523773 & 0.522247 & 0.509936\\
     1.6 & 0.505381 & 0.524250 & 0.522699 & 0.510186\\
     1.7 & 0.505513 & 0.524588 & 0.523020 & 0.510371\\
     1.8 & 0.505618 & 0.524835 & 0.523255 & 0.510512\\
     1.9 & 0.505701 & 0.525020 & 0.523432 & 0.510621\\
     2.0 & 0.505769 & 0.525163 & 0.523569 & 0.510709\\
     \hline
    \end{tabular}
\caption{Table shows the values of $E_g$ at different radial distances for the linear case.}
\label{tab:Eg_values}
\end{table}

\subsection{Volume Integral Quantifier}\label{sec-VIQ}
\begin{figure}[H]
\centering
    \subfigure[GR]{\includegraphics[scale=0.35]{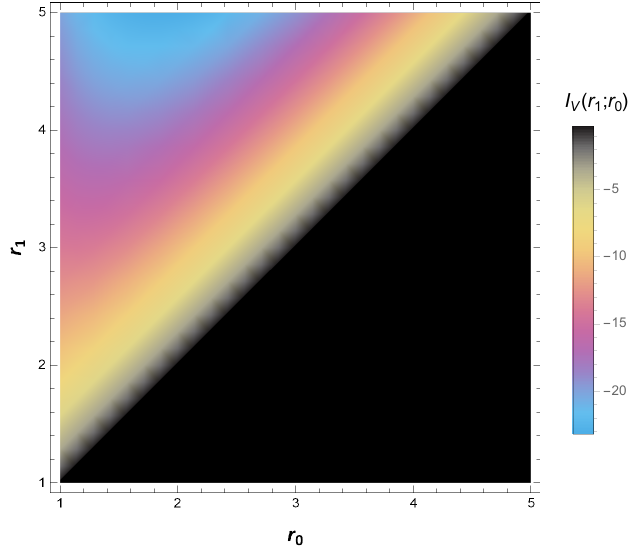}\label{}}
    \subfigure[$\alpha=2,\,\beta=0$]{\includegraphics[scale=0.35]{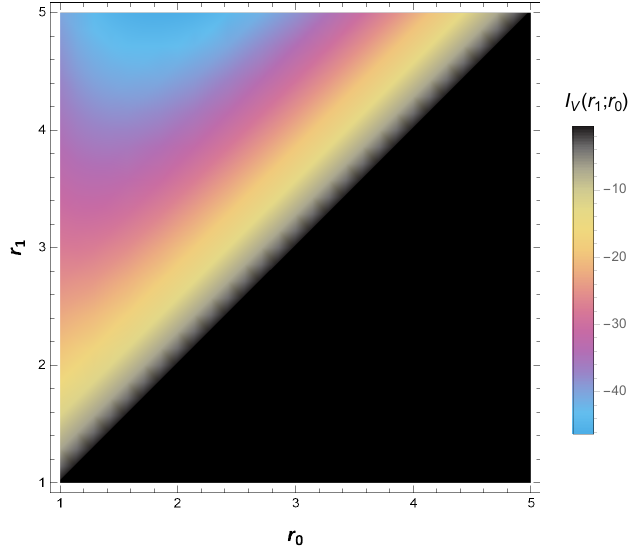}\label{}}
    \subfigure[$\alpha=2,\,\beta=-0.5$]{\includegraphics[scale=0.35]{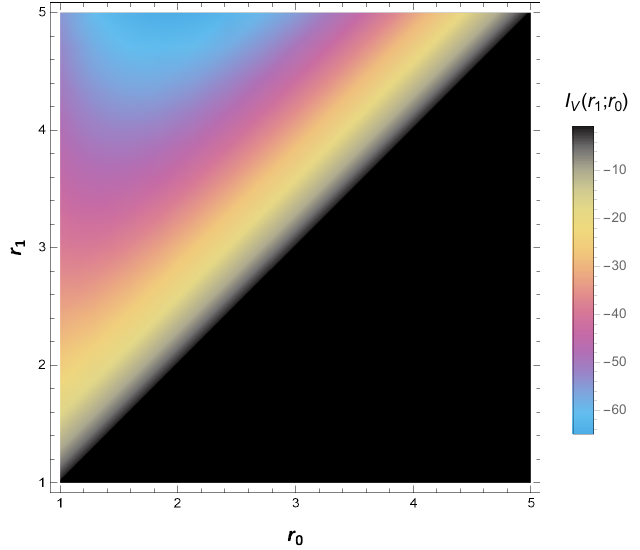}\label{}}
    \subfigure[$\eta=2.5,\,\chi=-0.5$]{\includegraphics[scale=0.3735]{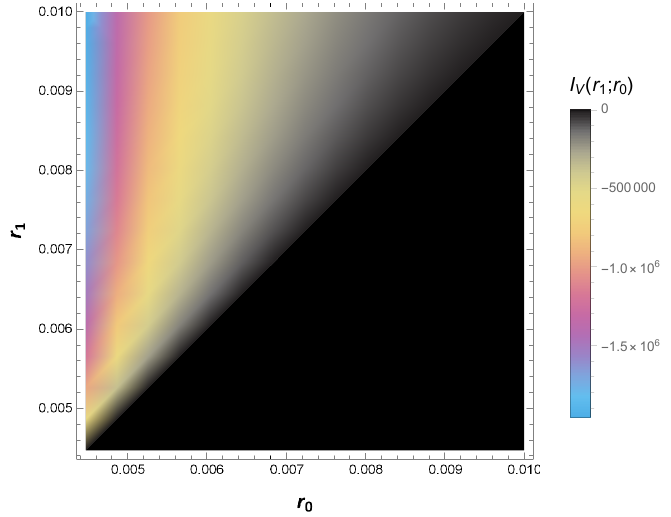}\label{}}
    \caption{Density Plots for the volume integral quantifier for the KMM correction with $\lambda=0.1$.}
    \label{fig:VIQdensity}
\end{figure}

The Volume Integral Quantifier (VIQ) informs us about the total amount of Averaged Null Energy Condition (ANEC) violating matter in spacetime \cite{visser2003traversable}. In effect, we know the ``total amount" of exotic matter required for the wormhole's maintenance. This is done by evaluating the following volume integral
\begin{align}
    I_V&=\oint[\rho(r)+p_r(r)]dV=2\int_{r_0}^\infty[\rho(r)+p_r(r)]dV
\end{align}
\begin{equation}
    \implies I_V=8\pi\int_{r_0}^\infty[\rho(r)+p_r(r)]r^2dr,
\end{equation}
where we have used $dV$ as $4\pi r^2dr$, the volume element in spherical coordinates.\\
We will evaluate the VIQ integral for our shape function $b(r)$, and it only makes sense to evaluate this integral at finite bounds, relevant to the wormhole's proximity. Therefore, we introduce a cut-off distance at $r_1>r_0$ to keep our expressions compact.
\begin{equation}\label{VIQ_integral}
I_V=8\pi\int_{r_0}^{r_1}[\rho(r)+p_r(r)]r^2dr.
\end{equation}
Upon evaluating the above integral using the shape function solution (\ref{simplified_b(r)}), we get
\begin{equation}
    I_V(r_1;r_0)=\frac{8\pi\alpha}{\beta+2}\Biggl[\frac{\gamma}{2}\left(\frac{1}{r_1}-\frac{1}{r_0}\right)+\frac{4\sigma_i\gamma}{3}\left(\frac{1}{{r_1}^3}-\frac{1}{{r_0}^3}\right)+\mathcal{M}\left(\frac{1}{{r_1}^\frac{\beta}{\beta-4}}-\frac{1}{{r_0}^\frac{\beta}{\beta-4}}\right)\Biggr].
\end{equation}
where,
$\gamma=-\frac{\pi^2(\beta-1)(\beta+2)}{180\alpha},$
$\sigma_i=\frac{5}{3}\frac{\lambda\Lambda_i}{(12-2\beta)},$
$\mathcal{M}=\frac{2\beta-4}{\beta}\left({r_0}^\frac{2\beta-4}{\beta-4}-\frac{\gamma}{4}{r_0}^\frac{4}{\beta-4}-\sigma_i\gamma {r_0}^\frac{12-2\beta}{\beta-4}\right).$
In Fig.\ref{fig:VIQdensity}, we have plotted the graphs for the above expression of VIQ for different throat radii and radial distance of the observer measuring the amount of ANEC-violating exotic matter. It is worth noting that $I_V$ is negative for all $r_1>r_0$. These small violations of the quantity can be attributed to quantum fluctuations. Similar results can be found in \cite{jusufi2020traversable}.
However, the non-linear case is an immediate outlier to the usual pattern of small traces of exotic matter. This is due to the non-linear nature of action integral and the exotic matter being concentrated at a small throat. If we were to detect high-frequency oscillations like in Table-\ref{tab:QNM_boson_quad} and \ref{tab:QNM_dirac}, apart from black holes, it could also be anticipated that such harmonic traces are given off by an exotic matter-rich wormhole.
\subsection{Equilibrium Analysis}\label{sec-Equil}
\begin{figure}[H]
\centering
    \subfigure[GR]{\includegraphics[scale=0.32]{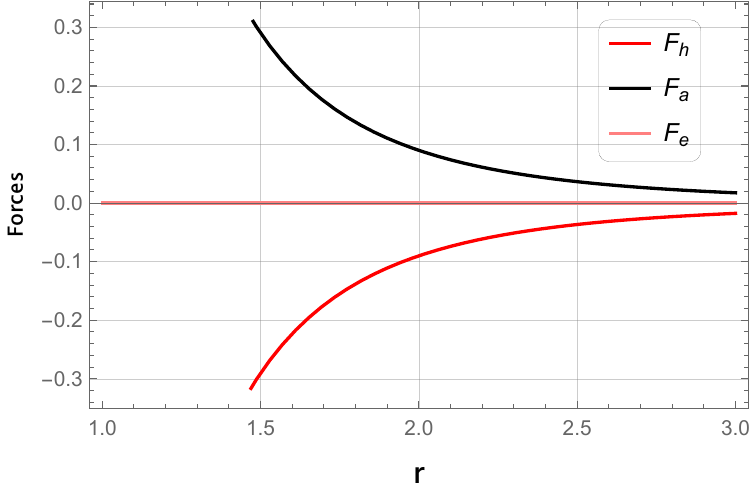}\label{}}\quad
    \subfigure[$\alpha=2,\,\beta=0$]{\includegraphics[scale=0.32]{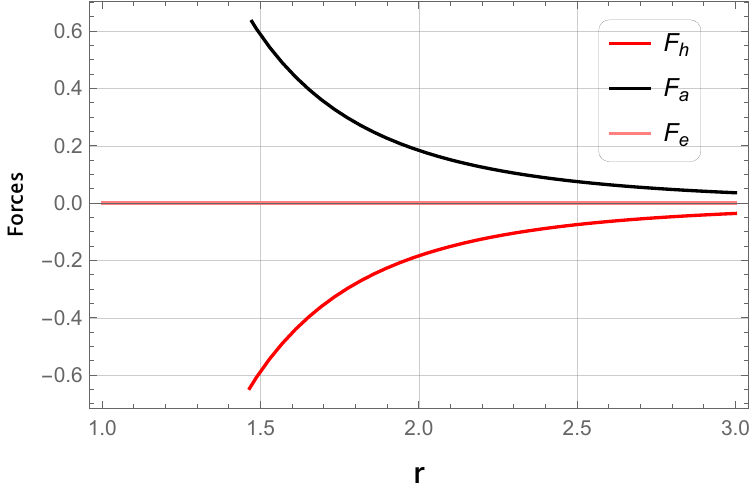}\label{}}\quad
    \subfigure[$\alpha=2,\,\beta=-0.5$]{\includegraphics[scale=0.32]{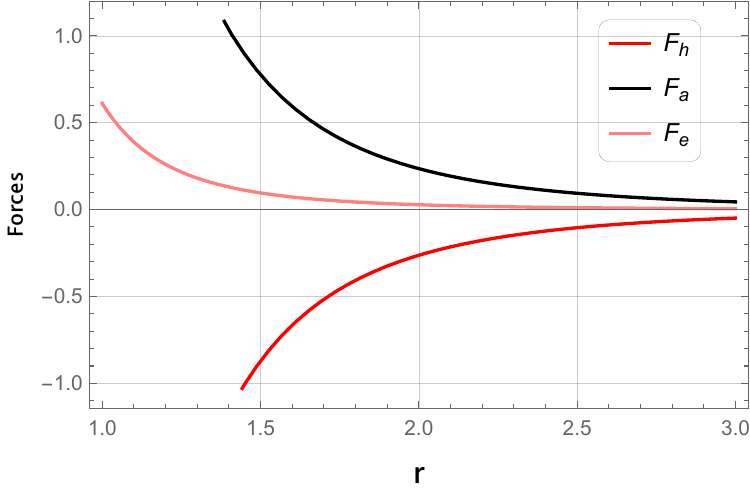}\label{}}\quad
    \subfigure[$\eta=2.5,\,\chi=-0.5$]{\includegraphics[scale=0.38]{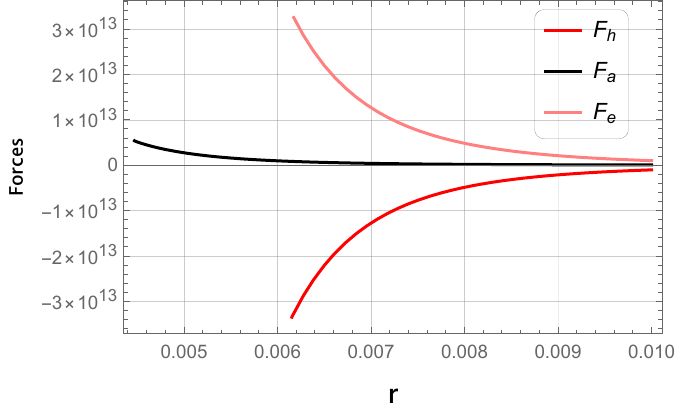}\label{}}
    \caption{Plot of the forces acting on the system for KMM corrections with $\lambda=0.1$.}
    \label{fig:Forces}
\end{figure}

In this section, we test the stability of the obtained wormhole solutions with the help of the generalized Tolman–Oppenheimer–Volkoff (TOV) equation. The generalized TOV equation can be defined as \cite{ditta2021study,mustafa2021traversable}
\begin{equation}\label{6d1}
-\frac{dp_r}{dr}-\frac{\nu'}{2}(\rho+p_r)+\frac{2}{r}\Delta-\frac{2\beta}{\beta+2}\Biggl(\frac{1}{4}\frac{d\rho}{dr}-\frac{1}{4}\frac{dp_r}{dr}-\frac{dp_t}{dr}\Biggl)=0.
\end{equation}
This, indeed, looks different from the conventional TOV equation due to the additional last term. It is a consequence of the matter coupling, which must be taken into consideration.
The above TOV equation \eqref{6d1} can be expressed in a much more elegant way as a sum of the forces in an equilibrium
\begin{equation}\label{tov}
    \mathcal{F}_h+\mathcal{F}_g+\mathcal{F}_a+\mathcal{F}_e=0,
\end{equation}
where
\begin{equation}
    \mathcal{F}_h=-\frac{dp_r}{dr},\;\;\;\mathcal{F}_g=-\frac{\nu'}{2}(\rho+p_r),\;\;\;\mathcal{F}_a=\frac{2}{r}(p_t-p_r),\;\;\;\mathcal{F}_e=-\frac{2\beta}{\beta+2}\Biggl[\frac{1}{4}\frac{d\rho}{dr}-\frac{1}{4}\frac{dp_r}{dr}-\frac{dp_t}{dr}\Biggl]
\end{equation}
are the hydrostatic, gravitational, anisotropic forces, and extra force, respectively. All the above-mentioned forces cancel each other out, effectively giving us an idea about the system's stability (in the classical sense unlike in Sec.\ref{sec-QNMs} which is mostly concerned with stability in the quantum level) and equilibrium. Since we are working in the absence of any tidal force, i.e. $\nu(r)=0$, the force $\mathcal{F}_g$ is zero.\\
For the linear case, we have used the shape function (\ref{simplified_b(r)}) and calculated the expressions of hydrostatic, anisotropic, and extra forces
\begin{align}\label{f1}
    \mathcal{F}_h=-\frac{\alpha}{(\beta-1)(\beta+2)r^4}\Biggl(&\notag\frac{8(\beta-1)(\beta-2)(\beta-3)}{(\beta-4)^2}\frac{{r_0}^\frac{2\beta-4}{\beta-4}}{r^\frac{\beta}{\beta-4}}+\gamma\Biggl\{\frac{2\beta-1}{r}-\frac{2(\beta-1)(\beta-2)(\beta-3)}{(\beta-4)^2}\frac{{r_0}^\frac{4}{\beta-4}}{r^\frac{\beta}{\beta-4}}\\
    &+\sigma_i\Biggl[\frac{21\beta-6}{r^3}-\frac{8(\beta-1)(\beta-2)(\beta-3)}{(\beta-4)^2}\frac{{r_0}^\frac{12-2\beta}{\beta-4}}{r^\frac{\beta}{\beta-4}}\Biggl]\Biggl\}\Biggr),
\end{align}
\begin{align}\label{f2}
    \mathcal{F}_a=\frac{\alpha}{(\beta+2)r^4}\Biggl(\left(\frac{4\beta-12}{\beta-4}\right)\frac{{r_0}^\frac{2\beta-4}{\beta-4}}{r^\frac{\beta}{\beta-4}}+\gamma\Biggl\{\frac{1}{r}+\left(\frac{\beta-3}{\beta-4}\right)\frac{{r_0}^\frac{4}{\beta-4}}{r^\frac{\beta}{\beta-4}}+\sigma_i\Biggl[\frac{6}{r^3}+\left(\frac{4\beta-12}{\beta-4}\right)\frac{{r_0}^\frac{12-2\beta}{\beta-4}}{r^\frac{\beta}{\beta-4}}\Biggl]\Biggl\}\Biggr),
\end{align}
\begin{align}\label{f3}
    \mathcal{F}_e=\frac{\alpha\beta}{(\beta-1)(\beta+2)r^4}\Biggl(\frac{4(\beta-1)(\beta-3)}{(\beta-4)^2}\frac{{r_0}^\frac{2\beta-4}{\beta-4}}{r^\frac{\beta}{\beta-4}}+\gamma\Biggl\{\frac{1}{r}-\frac{(\beta-1)(\beta-3)}{(\beta-4)^2}\frac{{r_0}^\frac{4}{\beta-4}}{r^\frac{\beta}{\beta-4}}+\sigma_i\Biggl[\frac{15}{r^3}-\frac{4(\beta-1)(\beta-3)}{(\beta-4)^2}\frac{{r_0}^\frac{12-2\beta}{\beta-4}}{r^\frac{\beta}{\beta-4}}\Biggl]\Biggl\}\Biggr).
\end{align}
As for the non-linear model, we resort to the use of numerical methods for the relevant calculations. Fig.\ref{fig:Forces} graphically presents the behavior of these forces in the neighborhood of the throat. Since the wormhole geometry in both cases displays positive pressure anisotropy $\Delta>0$ due to the presence of exotic matter, the anisotropic force is also positive. Likewise, the hydrostatic force is negative due to the increasing nature of $p_r(r)$. However, as one may expect, the extra forces only appear when the spacetime geometry is coupled with the energy and matter in it, i.e., for $\beta=0$, $\mathcal{F}_e$ is absent. This is a characteristic of geometry-matter-coupled gravity theories in general. Test objects experiencing this extra force deviate from the geodesic equation given by GR. It is mathematically established from the TOV equation that these forces will balance out each other in the bigger picture. Therefore, it is safe to state that the wormhole solutions derived in this paper for both models are classically stable. For further exploration of this subject, interested readers can refer to \cite{Sokoliuk}, where the authors have extensively explored this topic.

\section{Conclusions}\label{sec-conclusion}
In this paper, we have studied traversable wormholes under the effect of Casimir energy in the $f(T, \mathcal{T})$ gravity framework. The Casimir effect, which can be recreated in a laboratory apparatus, occurs due to the distortion of the quantum field of vacuum between two uncharged conducting plates, which act as a boundary condition for the quantum electrodynamics. This results in an attractive force between the plates occurring due to the negative zero-point energy, dubbed as the ``Casimir Energy". Such a form of energy may indicate the presence of exotic matter in an enclosed region of space, which is precisely the main ingredient for a traversable wormhole that violates the energy conditions at its throat. In this paper, we have studied the traversable wormhole solutions in the framework of $f(T, \mathcal{T})$ gravity by exploring the dynamics of the quantum nature of the Casimir effect. We have modified the Casimir energy with three GUP models, namely KMM, DGS, and Model II, to show the effect of GUP on the Casimir wormhole.  In Ref. \cite{jusufi2020traversable}, the authors explored the effect of GUP on Casimir wormholes by applying the aforementioned GUP relations in GR. Subsequently, the effect of KMM and DGS models on wormhole geometry has been discussed in $f(R, T)$ \cite{TRIPATHY2021100757} as well as $f(Q)$ gravity \cite{hassan2023gup}.\\
\indent We first considered the function $f(T, \mathcal{T})=\alpha T+\beta\mathcal{T}$ and obtained the analytic solution of shape function by comparing the density component of the field equations with the GUP-corrected Casimir energy density in Eq. \eqref{sol_const}. The resulting shape function behaves as expected for a wormhole geometry described by the Morris-Thorne metric, with certain restrictions on the parameters $\alpha$ and $\beta$. It is evident that the GUP parameter $\lambda$ significantly impacts the area outside the wormhole's throat. We also observed that as $\lambda$ increases, there is a substantial rise in the shape function within the wormhole throat. 
We then examined the energy conditions near the wormhole throat for each GUP model. It was found that $\rho+p_r\geq 0$ is violated near the throat, thus disobeying the NEC. This violation is desirable for a traversable wormhole, as it helps the wormhole geometry remain open rather than collapsing in on itself. 
Notably, the model parameters $\alpha$ and $\beta$ play a significant role in influencing all the energy conditions, especially the NEC. We also note that in violation of NEC, $f(T,\mathcal{T})$ gravity gives more contribution compared to the GR as well as $f(T)$ gravity. This is because matter coupling parameter $\beta$ plays an important role in the violation of energy conditions. The GUP correction to the Casimir wormholes also affects the energy conditions, with the NEC being violated beyond the wormhole throat, and this violation is maintained with the GUP correction.
Further, we examined the pressure anisotropy defined as $\Delta=p_t-p_r$, which showed positive behavior throughout the domain and confirmed the presence of exotic matter at the throat. Furthermore, EoS parameters were discussed, which shows diverging $r\rightarrow \infty$. Here, we noticed that modified gravity parameters significantly influence the behaviors of the EoS parameters, particularly when considered at distances away from the wormhole throat.\\
\indent Next, we have investigated the existence of wormhole solution for the non-linear model $f(T,\mathcal{T})=\eta T^2+\chi\mathcal{T}$ within the framework of each GUP model. We used numerical techniques to set an initial condition and studied the graphical behavior of shape functions, energy conditions, pressure anisotropy, and EoS parameters (refer to Figs. \eqref{fig:Shape function_quad} and \eqref{fig:EC_quad}). Our observations indicated that the shape function exhibits a positive increasing behavior near the throat and shows decreasing behavior at a significant distance from the throat. We also determined the throat radius to be approximately: $0.0045$ (KMM), $0.0037$ (DGS), and $0.0019$ (Model II). While we found that the flare-out condition is satisfied, the asymptotic condition is only met for a small radius near the throat due to the non-linearity of the Lagrangian, which is inevitable because of quantum corrections. This kind of behavior of the shape function can be found in \cite{hassan2023gup}. Additionally, we investigated the NEC and SEC in the vicinity of the throat and found that both were violated at the wormhole throat. Furthermore, the pressure anisotropy of the spacetime is strictly positive and projects a monotonically decreasing trend.\\
\indent The QNMs of the above wormhole solutions were studied for scalar, vector (electromagnetic), tensor (gravitational), and Dirac field perturbations. When such external field perturbations act around the throat, the wormhole starts to pulse out the additional energy, which travels as ripples in the spacetime manifold to regain its original state. These oscillations are characterized by their frequency which depends on the behaviour of the peak of the potential present at its throat. The equations of motion for aforementioned fields are reduced to a one-dimensional time-independent Schr\"odinger equation \eqref{red_wave_eq} in the tortoise coordinate, which governs the dynamics of the quasinormal states. The eigenfrequencies/QNM frequencies are found using the WKB approximation technique up to the third order. These frequencies are complex numbers whose real part gives the frequency of the oscillations, and the imaginary part tells us about the damping rate. One may refer to Tables-\ref{tab:QNM_boson_lin}, \ref{tab:QNM_boson_quad} and \ref{tab:QNM_dirac} to see the resulting frequencies. It is observed that the imaginary parts of the QNM frequencies are negative, which indicates that these oscillations are damped and the system will eventually regain its original state. Furthermore, the oscillation frequencies and damping are much larger for the non-linear model compared to the linear counterpart. This is yet again a consequence of the above-mentioned non-linearity of the Lagrangian and quantum corrections. This work has a lot of ``potential" to be studied further, starting with the time domain analysis for the quasinormal states or extending the WKB approximation up to higher-order terms to achieve greater accuracy in results.\\
\indent We discussed some important physical features and attributes of wormholes. As expected, our wormhole solutions possess a negative active mass due to the negative Casimir energy density. We also numerically calculated the total gravitational energy ($E_g$) using the Mathematica function `$NIntegrate$'. A table with values for $E_g$ at different radial distances `$r$' can be found in Table-\ref{tab:Eg_values}. We observed that this energy is positive near the throat $r_0$, which is a characteristic behavior of exotic matter, in contrast to the negative gravitational energy for regular baryonic matter. Additionally, we discussed VIQ analytically, which effectively tells us the amount of exotic matter required to violate the averaged null energy condition (ANEC). We can better understand the VIQ by looking at Fig. (\ref{fig:VIQdensity}), which shows the required amount of exotic matter for different sizes of the throat $r_0$ and the radial distance $r_1$ up to which the ANEC is violated. Considering that wormhole geometries may be unstable for non-traversable wormholes, we investigated the stability of our wormhole solution using the generalized TOV equation. We found that all the physical forces acting on the wormhole effectively cancel out, allowing the throat to remain open.\\
Based on current literature, research has been conducted on the dynamics of Casimir wormholes within the framework of  GR and in modified theories of gravity. R. Garattini \cite{garattini2019casimir} proposed the first formal solution for a traversable Casimir wormhole and examined the consequences of quantum weak energy conditions. Subsequently, Jusufi et al. \cite{jusufi2020traversable} studied the GUP correction to Casimir wormholes in GR. Alencar et al. \cite{alencar2021casimir} demonstrated that in $(2+1)$ dimensions, the Casimir energy density and pressure cannot support a traversable wormhole geometry, whereas traversable Casimir wormholes are supported in higher dimensional spacetimes with $D > 3$ \cite{oliveira2022traversable}. Santos et al. \cite{santos2023yang} investigated three-dimensional Casimir wormholes sourced by Casimir energy density and pressures related to the quantum vacuum fluctuations in Yang-Mills theory.
In \cite{TRIPATHY2021100757}, the author explored Casimir wormholes with KMM and DGS models in the context of $f(R,T)$ gravity. Later, using the same KMM and DGS model, Casimir wormholes have been explored in different modified theories of gravity, such as in $f(Q)$ gravity \cite{hassan2023gup}, $f(Q,T)$ gravity \cite{sahoo2024casimir} and EGB gravity \cite{zubair2023imprints}. Additionally, Casimir wormholes with KMM and DGS models in $f(R,T)$ gravity with conformal killing vector \cite{Mushayydha2023} and in EGB gravity \cite{FAROOQ2023169542} with non-commutative geometries have been studied. In this paper, we have considered GUP Model-II in addition to the KMM and DGS model for the study of the Casimir wormholes in $f(T,\mathcal{T})$ gravity. We present a detailed analysis of the effects of the GUP on the wormhole geometry that sources its exotic matter from the Casimir energy in the setting of torsion-matter coupled $f(T,\mathcal{T})$ gravity. The exact analytical solution to $n^{th}$-order GUP corrections to the Casimir energy is provided, which can be utilized for future research when further correction terms are discovered. We also discussed the deviation of behavior in the wormhole solutions by comparing them with GR and $f(T)$ gravity. 
Additionally, we explored the QNMs of the Casimir wormhole, which has not been discussed yet in any modified theories of gravity.
Note that the GUP correction up to the first order on the minimal length scale has been applied, and this approach can also be used to compute the corrections of Casimir energy up to the next leading order and investigate its significance on the wormhole geometry \cite{Nouicer_2005,panella2007casimir}.

\acknowledgments
ZH acknowledges the Department of Science and Technology (DST), Government of India, New Delhi, for awarding a Senior Research Fellowship (File No. DST/INSPIRE Fellowship/2019/IF190911). PKS acknowledges the National Board for Higher Mathematics (NBHM) under the Department of Atomic Energy (DAE), Govt. of India, for financial support to carry out the Research project No.: 02011/3/2022 NBHM(R.P.)/R\&D II/2152 Dt.14.02.2022. A. {\"O}. would like to acknowledge the contribution of the COST Action CA21106 - COSMIC WISPers in the Dark Universe: Theory, astrophysics and experiments (CosmicWISPers) and the COST Action CA22113 - Fundamental challenges in theoretical physics (THEORY-CHALLENGES). We also thank TUBITAK and SCOAP3 for their support.

\bibliography{ref.bib}

\end{document}